\documentclass[aps, pre, onecolumn, nofootinbib, notitlepage, groupedaddress, amsfonts, amssymb, amsmath, longbibliography, superscriptaddress]{revtex4-1}
\usepackage{tabularx}
\usepackage{graphicx}
\usepackage{hyperref}
\usepackage{xcolor}
\usepackage{placeins}
\hypersetup{
    colorlinks,
    linkcolor={red!50!black},
    citecolor={blue!50!black},
    urlcolor={blue!80!black}
}
\usepackage{bm}
\usepackage{natbib}
\usepackage{longtable}
\newcommand{\DivU}{\ensuremath{\nabla\cdot\bm{u}}}
\newcommand{\angles}[1]{\ensuremath{\left\langle #1 \right\rangle}}

\newcommand{\grad}{\ensuremath{\nabla}}
\newcommand{\RB}{Rayleigh-B\'{e}nard }
\newcommand{\Reff}{\ensuremath{\text{Re}_{\text{ff}}}}
\newcommand{\Peff}{\ensuremath{\text{Pe}_{\text{ff}}}}

\newcommand\mnras{{MNRAS}}%
\newcommand\apjl{{The Astrophysical Journal Letters}}%

\begin{document}
\author{Evan H. Anders}
\affiliation{Dept. Astrophysical \& Planetary Sciences, University of Colorado -- Boulder, Boulder, CO 80309, USA}
\affiliation{Laboratory for Atmospheric and Space Physics, Boulder, CO 80303, USA}
\author{Geoffrey M. Vasil}
\affiliation{University of Sydney School of Mathematics and Statistics, Sydney, NSW 2006, Australia}
\author{Benjamin P. Brown}
\affiliation{Dept. Astrophysical \& Planetary Sciences, University of Colorado -- Boulder, Boulder, CO 80309, USA}
\affiliation{Laboratory for Atmospheric and Space Physics, Boulder, CO 80303, USA}
\author{Lydia Korre}
\affiliation{Laboratory for Atmospheric and Space Physics, Boulder, CO 80303, USA}

\title{Convective dynamics with mixed temperature boundary conditions: \\
why thermal relaxation matters and how to accelerate it}

\begin{abstract}
Astrophysical simulations of convection frequently impose different thermal boundary conditions at the top and the bottom of the domain in an effort to more accurately model natural systems.
In this work, we study \RB convection (RBC) under the Boussinesq approximation.
We examine simulations with mixed temperature boundary conditions in which the flux is fixed at the bottom boundary and the temperature is fixed at the top (``FT'').
We aim to understand how FT boundaries change the nature of the convective solution compared to the traditional choice of thermal boundaries, in which the temperature is fixed at the top and bottom of the domain (``TT'').
We demonstrate that the timescale of thermal relaxation for FT simulations is dependent upon the initial conditions.
``Classic'' initial conditions which employ a hydrostatically- and thermally- balanced linear temperature profile exhibit a long thermal relaxation.
This long relaxation is not seen in FT simulations which use a TT simulation's nonlinear state as initial conditions (``TT-to-FT'').
In the thermally relaxed, statistically stationary state, the mean behavior of an FT simulation corresponds to an equivalent simulation with TT boundaries, and time- and volume-averaged flow statistics like the Nusselt number and the P\'{e}clet number are indistinguishable between FT and TT simulations.
FT boundaries are fundamentally asymmetric, and we examine the asymmetries that these boundaries produce in the flow.
We find that the fixed-flux boundary produces more extreme temperature events than the fixed-temperature boundary.
However, these near-boundary asymmetries do not measurably break the symmetry in the convective interior.
We briefly explore rotating RBC to demonstrate that our findings with respect to thermal relaxation carry over to this more complex case, and to show the power of TT-to-FT initial conditions.
\end{abstract}
\maketitle


\section{Introduction}
\label{sec:introduction}
Convection is a crucial heat transport mechanism in the atmospheres and interiors of stars and planets.
Numerical simulations are a commonly-used tool in studies of geophysical or astrophysical convection.
These studies range from examinations of convection in the simplified Boussinesq approximation \cite{spiegel&veronis1960, ahlers&all2009, plumley&julien2019} to highly complex ``dynamo simulations'' which include magnetism and atmospheric density stratification \cite{charbonneau2014, toomre2019}.
Regardless of complexity, numerically simulated convection is fundamentally driven by some combination of imposed boundary conditions and internal heating profiles \cite{goluskin2016}.
In studies of Boussinesq convection, the standard choice is to hold constant the temperature difference across the domain by fixing the temperature at the upper and lower boundaries.
However, a common choice of thermal boundary conditions in astrophysical convection \cite{glatzmaier&gilman1982, hurlburt&all1986, cattaneo&all1990, featherstone&hindman2016a, korre&all2019, wood&brummell2018, kapyla&all2019, matilsky&all2019} is to fix the flux entering the domain through the bottom boundary and to fix the value of a thermodynamic quantity (e.g., temperature or entropy) at the top boundary.
We are unaware of any study which has examined the consequences of imposing these ``mixed'' boundaries that are frequently favored in astrophysical convection studies.

In this work, we examine how the choice of using ``mixed'' thermodynamic boundary conditions affects the evolved nonlinear convective state in the simplest possible model: \RB convection (RBC) under the Boussinesq approximation.
In RBC, temperature is the only thermodynamic quantity and throughout this work we will adopt the gnomenclature of past authors (see e.g., ref.~\cite{ishiwatari&all1994}) and refer to the choice of fixing the flux at the bottom and temperature at the top as ``FT'' boundary conditions.
We will refer to the common choice of fixing temperature at both boundaries as ``TT'' boundaries, and fixing the flux at both boundaries as ``FF'' boundaries\footnote{Note, in ref.~\cite{goluskin2016}, our TT, FF, and FT are respectively called RB1, RB2, and RB3.}.

It is generally assumed that, in their statistically stationary states, simulations with FT boundaries should behave similarly to those with FF boundaries \cite{goluskin2016, otero&all2002}.
Early studies of FF convection often focused on flow morphologies, because large-to-infinite aspect ratio convective rolls are linearly unstable for this choice of boundary condition (see e.g., ref.~\cite{chapman&proctor1980}).
However, the onset properties and resultant flow morphologies in FT simulations more strongly resemble TT dynamics \cite{ishiwatari&all1994}, in that both are linearly unstable at a well-defined, finite aspect ratio.
Despite these differences near convective onset, FF and TT boundaries have been shown to exhibit the same scaling of convective heat transport (quantified by the Nusselt number, Nu) as a function of increased convective driving (quantified by the Rayleigh number, Ra) \cite{johnston&doering2009}.
However, FT boundaries introduce complexities into the convective solution which neither FF nor TT boundaries are exposed to.
First, the evolved mean temperature of a simulation with FT boundaries differs from the initial mean temperature, and therefore the thermal reservoir of the convective system must evolve (``thermally relax'') over time \cite{anders&all2018}.
Second, FT boundary conditions are fundamentally asymmetric, and it is unclear if these asymmetries affect the evolved convective solution.

In this paper, we investigate the thermal relaxation of, and the asymmetries in, RBC with FT boundary conditions.
We also compare relaxed FT solutions to TT solutions.
We find that when classic initial conditions which are in hydrostatic and thermal equilibrium are employed, the thermal relaxation of FT systems is very long compared to TT systems, in which it is nearly instantaneous.
The thermal relaxation of FT simulations is analogous to a sweep through parameter space in which dynamics are sampled over a range of values of Ra.
We also find that this long thermal relaxation can be bypassed by constructing smarter initial conditions based on the expected evolved value of Nu, or by simply using the results of TT simulations as initial conditions for FT simulations.
Finally, FT boundaries create some asymmetries in the convective flows, particularly in the boundary layers, but these asymmetries do not appreciably change the bulk convective state compared to TT simulations.

We present these findings as follows.
In section \ref{sec:simulations}, we describe our simulation setup, numerical methods, initial conditions, and timescales in the convective systems.
In section \ref{sec:results_timescales}, we describe our findings regarding the time evolution of FT systems.
In section \ref{sec:results_dynamics}, we study asymmetries in FT systems and compare them to TT systems.
In section \ref{sec:results_rotating}, we show that these findings carry over to a more complex system (rotating \RB convection) with some interesting implications.
Finally, in section \ref{sec:discussion}, we summarize our findings and briefly describe the implications of this work for the field of astrophysical convection.

\section{Simulation Details}
\label{sec:simulations}

\subsection{Equations, Control Parameters, Boundary Conditions, and Numerics}
We study incompressible RBC under a freefall nondimensionalization; for details of this nondimensionalization, we refer readers to our previous work \cite{anders&all2018}.
In section \ref{sec:results_rotating}, we study convection in the presence of vertical global rotation \cite{julien&all1996}, and include the Coriolis term in the momentum equation for generality.
The Boussinesq equations of motion are
\begin{align}
\DivU &= 0
	\label{eqn:incompressible}
\\
\frac{\partial \bm{u}}{\partial t} + \left(\bm{\omega} + \frac{1}{\text{Ek }\Reff}\hat{z}\right)\times\bm{u} 
&= - \grad \varpi + T_1\hat{z} - \frac{1}{\Reff}\grad\times\bm{\omega},
	\label{eqn:bouss_momentum}
\\
\frac{\partial T_1}{\partial t}  + \bm{u}\cdot\grad T_1 + w \frac{\partial T_0}{\partial z} 
&= \frac{1}{\Peff}\grad^2 T_1,
	\label{eqn:bouss_energy}
\end{align}
where $\bm{u} = (u, v, w)$ is the velocity, $T = T_0(z) + T_1(x, y, z, t)$ is the temperature (where $T_0$ is a background linearly unstable temperature profile and $T_1$ are the fluctuations around that profile), $\varpi$ is the reduced kinematic pressure \cite{anders&all2018} which enforces the incompressibility constraint, and $\bm{\omega} = \grad \times \bm{u}$ is the vorticity.
The dimensionless control parameters are the Rayleigh (Ra), Prandtl (Pr), and Ekman (Ek) numbers, defined respectively as
\begin{equation}
\text{Ra} = \frac{g \alpha L_z^3 \Delta}{\nu\kappa} = \frac{(L_z\,u_{\text{ff}})^2}{\nu\kappa}, \qquad \text{Pr} = \frac{\nu}{\kappa}, \qquad \text{Ek} = \frac{\nu}{2\Omega L_z^2},
\end{equation}
where $u_{\text{ff}}$ is the freefall velocity, $g$ is the gravitational acceleration, $\alpha$ is the coefficient of thermal expansion, $L_z$ is the domain depth, $\nu$ and $\kappa$ are respectively the viscous and thermal diffusivity, $\Omega$ is the global rotation frequency, and $\Delta$ is the nondimensional temperature scale (defined below).
These parameters set the freefall Reynolds (\Reff) and P\'{e}clet (\Peff) numbers, 
\begin{equation}
\Reff = \sqrt{\frac{\text{Ra}}{\text{Pr}}},\qquad
\Peff = \text{Pr }\Reff,
\end{equation}
and throughout this work we hold Pr = 1 so that $\Reff = \Peff$.
In non-rotating RBC (sections \ref{sec:results_timescales} \& \ref{sec:results_dynamics}), we set Ek$\,= \infty$.

The extent of our numerical domain vertically is $z = [-0.5, 0.5]$ and horizontally is $x, y = [-\Gamma/2, \Gamma/2]$, where $\Gamma$ is the aspect ratio.
The background temperature profile, $T_0(z) = 0.5 - z$, is unstable and linearly decreases from a value of 1 to 0 across the domain. 
The temperature scale, $\Delta$, is set by either the temperature jump across the domain ($\Delta = \Delta T_0 =  T_0(z=0.5)-T_0(z=-0.5)$) for TT boundaries or by the temperature gradient lengthscale ($\Delta = L_z \partial_z T_0$) for FT boundaries.
We respectively define a temperature (Ra$_{\Delta T}$) and a flux (Ra$_{\partial_z T}$) Rayleigh number for these cases,
\begin{equation}
\text{Ra}_{\Delta T} = \frac{g \alpha L_z^3 \Delta T_0}{\nu\kappa}, \qquad 
\text{Ra}_{\partial_z T} = \frac{g \alpha L_z^4 \partial_z T_0}{\nu\kappa}.
\end{equation}
We respectively impose TT and FT boundary conditions as
\begin{equation}
(\text{TT}): T_1 = 0 \text{ at $z$ = \{-0.5, 0.5\}},\qquad\qquad
(\text{FT}): \partial_z T_1 = 0 \text{ at $z$ = -0.5} \,\,\,\&\,\,\, T_1 = 0 \text{ at $z$ = 0.5}.
\end{equation}

In sections \ref{sec:results_timescales} \& \ref{sec:results_dynamics}, we study non-rotating convection.
For comparison with the literature, we specify $\Gamma = 2$ and these simulations employ no-slip, impenetrable boundaries,
\begin{equation}
u = v =\, w = 0 \, \, \text{at}\,\,z = \{-0.5, 0.5\}.
\label{eqn:vel_bcs}
\end{equation}
For this choice of boundary conditions, the critical values of the Rayleigh number and wavenumber are (Ra$_{\partial_z T}, k) = (1295.78, 2.5519)$ for FT boundaries and (Ra$_{\Delta T}, k) = (1707.76, 3.1163)$ for TT boundaries \cite{goluskin2016}.
In our $\Gamma = 2$ box, the smallest wavenumber permitted is $k = \pi$, and at that wavenumber the critical values are Ra$_{\partial_z T} = 1357.57$ for FT boundaries and Ra$_{\Delta T} = 1707.94$ for TT boundaries, which are slightly larger than the classical onset values.
It is reasonable to expect important differences between FT and TT solutions at low supercriticalities due to the difference in onset.
However, for the supercriticalities of O($10^{5+}$) studied here, we do not expect this difference in linear stability to be very important.
Many of these simulations are restricted to two-dimensional (2D) convection by setting $\partial_y = v = 0$.

The rotating cases in section \ref{sec:results_rotating} employ stress-free, impenetrable boundaries,
\begin{equation}
\partial_z u = \partial_z v = w = 0 \, \, \text{at}\,\,z = \{-0.5, 0.5\}.
\label{eqn:vel_bcs}
\end{equation}
We follow previous work  \cite{stellmach&all2014} and study three-dimensional (3D) tall, skinny boxes with $\Gamma = 10\lambda_c(\text{Ek})$, where $\lambda_c(\text{Ek})$ is the wavelength of convective onset at the specified value of Ek. 
For the cases studied here at Ek = $10^{-6}$, and for TT boundaries, $\lambda_c(10^{-6}) \approx 4.81 \times 10^{-2}$ and the critical Rayleigh number is Ra$_{\Delta T} \approx 9.2 \times 10^{8}$.

We utilize the Dedalus\footnote{\url{http://dedalus-project.org/}} pseudospectral framework \cite{burns&all2016, burns&all2020} to evolve Eqs.~(\ref{eqn:incompressible}-\ref{eqn:bouss_energy}) forward in time.
Our 2D simulations use an implicit-explicit (IMEX), third-order, four-stage Runge-Kutta timestepping scheme RK443; our 3D simulations use the IMEX, second-order, two-stage Runge-Kutta scheme RK222 \cite{ascher&all1997}. 
Variables are time-evolved on a dealiased Chebyshev (vertical) and Fourier (horizontal, periodic) domain in which the physical grid dimensions are 3/2 the size of the coefficient grid.  
The codes used to run the simulations and to create the figures in this work are available publicly online in a repository of supplemental materials \cite{anders&all2020a_supp}\footnote{Simulations were conducted using v1.1.0 (for Nu-based ICs and 3D non-rotating simulations) and v1.0.1 (for all other cases) of our \texttt{boussinesq\_convection} github repository \cite{code:boussinesq_convection}.}.

\subsection{Output Quantities \& Mapping Between Temperature Nondimensionalizations}
\label{sec:ra_nu_relations}
Throughout this work we will measure and report the evolved value of the Nusselt number (Nu).
We define and measure Nu instantaneously as
\begin{equation}
\text{Nu} \equiv \angles{\frac{w T - \Peff^{-1} \partial_z T}{-\Peff^{-1} \angles{\partial_z T}}}
= 1 + \Peff\frac{\angles{w T}}{-\Delta T},
\end{equation}
where $\angles{}$ represent a volume average ($\angles{s} \equiv \iint s\,dx\,dz / \Gamma$ in 2D and $\angles{s} \equiv \iiint s\,dx\,dy\,dz / \Gamma^2$ in 3D for some scalar quantity $s$), and $\Delta T = \angles{\partial_z T}$ is the (negative) temperature difference between the top and bottom plate.
In a thermally relaxed, statistically stationary state \cite{calkins&all2015}, 
\begin{equation}
\text{Nu} = \frac{\beta L_z}{\Delta T}, \,\,\text{where}\,\,
\beta \begin{cases}
< -1 & \text{(TT)} \\
= -1 & \text{(FT)},
\end{cases},
\qquad
\Delta T \begin{cases}
= -1 & \text{(TT)} \\
\in [-1, 0) & \text{(FT)}
\end{cases},
\label{eqn:evolved_nu}
\end{equation}
and where $\beta$ is the temperature gradient achieved at the domain boundaries.
Nu is therefore the conversion between a temperature and flux nondimensionalization such that the thermally relaxed state of any convective solution is characterized by both a Ra$_{\Delta T}$ and Ra$_{\partial_z T}$ according to
\begin{equation}
\text{Ra}_{\partial_z T} =\text{Nu}\, \text{Ra}_{\Delta T},
\qquad
T_{\text{TT}} = \text{Nu}\, T_{\text{FT}},
\qquad
\bm{u}_{\text{TT}} = \sqrt{\text{Nu}}\, \bm{u}_{\text{FT}}.
\label{eqn:ra_relation}
\end{equation}
This mapping is presented by ref.~\cite{calkins&all2015} for a diffusion timescale nondimensionalization, and we have expanded it here for a freefall timescale nondimensionalization.

Throughout this work, we will also measure the evolved P\'{e}clet number (Pe) and in section \ref{sec:results_rotating} we will measure the Rossby number (Ro).
These nondimensional quantities are defined as
\begin{equation}
\text{Pe} = \angles{|\bm{u}|}\Peff,\qquad \text{Ro} = \angles{|\bm{\omega}|}\text{Ek }\Reff,
\end{equation}
where $|\bm{q}|$ represents the magnitude of the vector quantity $\bm{q}$.

\subsection{Initial Conditions}

\subsubsection{Temperature Initial Conditions}

In FT simulations, the evolved, nonlinear dynamics determine the magnitude of $\Delta T$.
As a result, the time evolution of FT simulations is sensitive to how well the initial conditions guess the evolved $\Delta T$.
For this reason, we will study FT simulations which employ three different initial states.

\paragraph{Classic ICs} 
Our first set of initial conditions are the ``classic'' initial conditions on which the system was nondimensionalized,
\begin{equation}
T_c(z) = T_0(z) = 0.5 - z.
\end{equation}

\paragraph{TT-to-FT}
As Eqn.~\ref{eqn:ra_relation} suggests, and as we will show in section \ref{sec:results_timescales}, the evolved state of each FT simulation corresponds to an equivalent TT simulation.
As a result, we will examine ``TT-to-FT'' initial conditions, in which we run a TT simulation through its convective transient to statistical equilibrium, then use the full evolved nonlinear state as initial conditions for an FT simulation.
To achieve this, we perform these steps:
\begin{enumerate}
\item Run a TT simulation to its statistically stationary state ($\sim100+$ freefall time units). 
Measure $\text{Nu}$ in that state.
\item Re-nondimensionalize from TT to FT according to Eqn.~\ref{eqn:ra_relation}.
\item Restart the simulation with FT boundaries and continue timestepping.
\end{enumerate}

\paragraph{Nu-based ICs} 
The similarity of TT and FT simulations in the statistically stationary state suggests that Nu~vs.~Ra scaling laws derived for TT simulations can be expected to hold for FT simulations.
According to Eqn.~\ref{eqn:ra_relation}, we can rearrange a given scaling law,
\begin{equation}
\text{Nu} = A \text{Ra}_{\Delta T}^{\alpha} \,\,\Rightarrow\,\, \text{Nu} = (A \text{Ra}_{\partial_z T}^{\alpha})^{1/(1+\alpha)},
\end{equation}
and use this law along with Eqn.~\ref{eqn:evolved_nu} to predict the evolved temperature jump in an FT simulation,
\begin{equation}
\Delta T = \frac{\beta L_z}{\text{Nu}} = -(A \text{Ra}_{\partial_z T}^{\alpha})^{-1/(1+\alpha)}.
\end{equation}
Our ``Nusselt-based'' initial conditions construct an initial temperature profile which is consistent with the bottom fixed-flux boundary condition but whose initial $\Delta T$ is determined by a specific Nu~vs.~Ra scaling law.
The vertical temperature derivative is,
\begin{equation}
\frac{\partial T_N}{\partial z} = (\grad T)_{\text{interior}} - \xi(z)[1 + (\grad T)_{\text{interior}}].
\label{eqn:nu_based_gradT}
\end{equation}
We set the initial temperature field by integrating Eqn.~\ref{eqn:nu_based_gradT} according to the top (fixed-temperature) boundary condition.
Discontinuous profiles are unstable in our spectral methods, so we utilize a smooth windowing function, $\xi(z)$, to set the temperature gradient to $-1$ near the boundaries,
\begin{equation}
\xi(z) = 1 + \frac{1}{2}\left(\text{erf}\left[\frac{z - (0.5 - 2\delta_\xi)}{0.5\delta_\xi}\right] - \text{erf}\left[\frac{z - (-0.5 + 2\delta_\xi)}{0.5\delta_\xi}\right]\right).
\end{equation}
Here, $\delta_\xi = -\Delta T / 2$ is an estimate of the boundary layer width and the temperature gradient in the interior of the domain, $(\grad T)_{\text{interior}}$, is determined by setting $\int \partial_z T_N dz = \Delta T$.
In this work, we use the best-fit law of ref.~\cite{johnston&doering2009} with $A = 0.138$ and $\alpha = 0.285$ when constructing Nu-based ICs.

\subsubsection{Additional Initial Conditions}
In all cases, we modify the initial temperature profile by specifying the value of $T_1$, rather than through modifications to the linearly unstable reference profile, $T_0$.
We furthermore assume that the initial temperature profile is in hydrostatic equilibrium, and solve for $\varpi$ accordingly.
We assume zero velocity in the initial state, except in the case of TT-to-FT simulations, where velocities are taken directly from the corresponding TT simulation and scaled according to Eqn.~\ref{eqn:ra_relation}.
For classic and Nu-based ICs,
we fill $T_1$ with random white noise whose magnitude is $10^{-6}/\Peff$, and which is vertically tapered to zero at the boundaries.
We filter this noise spectrum in coefficient space, such that only the lower 25\% of the coefficients have power; this low-pass filter is used to avoid populating the highest wavenumbers with noise in order to improve the stability of our spectral timestepping methods.

\subsection{Timescales}
\label{sec:timescales}
One result of the mapping in Eqn.~\ref{eqn:ra_relation} is that the nondimensional dynamical freefall timescale is a poor description of the \emph{evolved} freefall timescale.
The velocities in an FT simulation are smaller than the velocities in a TT simulation by a factor of $\sqrt{\text{Nu}}$.
As a result, every nondimensional simulation freefall time unit in a TT simulation samples a factor of $\sqrt{\text{Nu}}$ more dynamics than a time unit in an FT simulation.
We therefore define the \emph{evolved} freefall time, $\tau_{\text{ff, ev}} = \sqrt{\text{Nu}}$ (for FT simulations) or $\tau_{\text{ff, ev}} = 1$ (for TT simulations).
To ensure accurate comparisons, we measure flow statistics over multiples of $\tau_{\text{ff, ev}}$ rather than over multiples of the nondimensional time units.

The thermal energy reservoir of our 3D Cartesian systems and the rate of change of this reservoir due to conduction at the boundaries are derived from volume-integrals of the temperature and vertical fluxes,
\begin{equation}
E = \iiint_V T\,dx\,dy\,dz = \Gamma^2 \int_{z=-0.5}^{z=0.5} T\,dz,
\qquad
\frac{dE}{dt} = -\iiint_V \grad\cdot \bm{F}_{\text{cond}}\,dx\,dy\,dz = -\Gamma^2 F_{\text{cond,z}}\bigg|_{z=-0.5}^{z=0.5}
\end{equation}
where $\bm{F}_{\text{cond}} = -\Peff^{-1}\grad T$ is the conductive flux whose z-component is $F_{\text{cond,z}}$, and where $\int T dz = \Delta T / 2$ in RBC.
The thermal relaxation time of an RBC experiment is therefore
\begin{equation}
\tau_{\text{th}} = \frac{\Delta E}{dE/dt},
\,\,\text{with}\,\,
\Delta E = E(t=\infty) - E(t=0).
\end{equation}
In the case of TT boundary conditions, we expect $\Delta E = 0$, as the initial and final state have the same $\Delta T$.
The goal of our ``Nu-based'' and TT-to-FT initial conditions is to create FT systems in which $\Delta E \approx 0$, creating a system with a negligible relaxational timescale.

For the case of classic ICs in an FT system, $\Delta T(t=0) = -1$ and $\Delta T(t=\infty) = -\text{Nu}^{-1}$, and the temperature gradient at the bottom boundary is fixed at a value of -1.
The thermal relaxation timescale is therefore
\begin{equation}
\tau_{\text{th, FT-classic}} = \Peff \frac{\text{Nu}^{-1} - 1}{\partial_z T(z=0.5) + 1} \sim \frac{\sqrt{\text{Ra Pr}}}{|\partial_z T(z = 0.5)| - 1},
\end{equation}
where the final expression is for the large Ra case where Nu$^{-1} \ll 1$.
The magnitude of the temperature derivative achieved at the top boundary is initially very large ($|\partial_z T(z=0.5)|_{\text{early}} > \text{Nu}(t=\infty)$; see e.g., Fig.~1b in ref.~\cite{anders&all2018}), but decreases in magnitude throughout the evolution of a simulation, making it difficult to estimate the true thermal relaxation time.
However, it is reasonable to assume that the evolutionary timescale lies within the window $\sqrt{\text{Ra Pr}} \,\text{Nu}^{-1} \lesssim \tau_{\text{th, FT-classic}} \lesssim \sqrt{\text{Ra Pr}}$.
In practice, we find that our non-rotating simulations equilibrate in $\sim 2.5\sqrt{\text{Ra}\,\text{Pr}}/\text{Nu}$ nondimensional time units (see section \ref{sec:time_evolution_classic_ICs}), and our rotating simulation equilibrates in $\sim (2/3)\sqrt{\text{Ra}\,\text{Pr}}/\text{Nu}$ time units (see section \ref{sec:results_rotating}).

\section{Results: How initial conditions influence evolutionary timescales}
\label{sec:results_timescales}

\subsection{Classic initial conditions: long thermal relaxation}
\label{sec:time_evolution_classic_ICs}

\begin{figure}[p!]
\includegraphics[width=\textwidth]{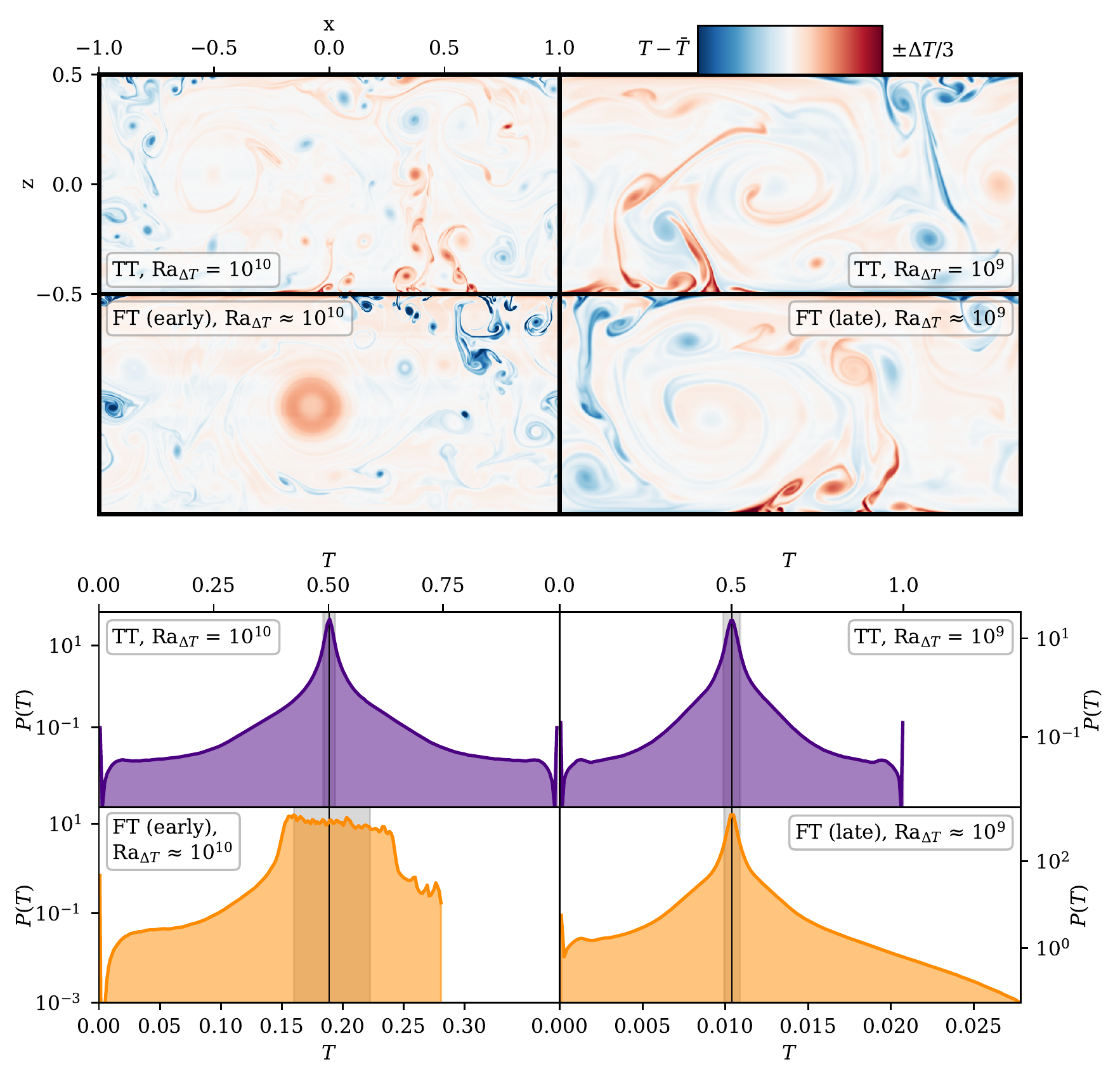}
\caption{ 
	(Upper four panels) Snapshots of the temperature anomaly in two TT simulations (top row) and in an FT simulation with Ra$_{\partial_z T} = 4.83 \times 10^{10}$ at early and late times (bottom row).
	(Left two panels) Dynamics in a TT case at Ra$_{\Delta T} = 10^{10}$ and early in the FT simulation when Ra$_{\Delta T} \approx 10^{10}$.
	To first order, both cases have similar flow structures: a large convective cell and plumes which break apart into small turbulent eddies.
	However, in the FT case, the temperature anomaly of the cold plume is much larger than the hot plume, which does not appear on this color scale.
	(Right two panels) Dynamics in a TT case at Ra$_{\Delta T} = 10^9$ and in the relaxed state of the previously pictured FT simulation with Ra$_{\Delta T} \approx 10^9$.
	The relaxed FT simulation is visually indistinguishable from its comparable TT simulation.
	Note that Ra$_{\Delta T}$ is a measured \emph{output} quantity for the FT simulation at two different times in its evolution.
\\	
	(Bottom four panels) Probability distribution functions (PDFs) of the full temperature field in each of the four dynamical panels pictured above.
	The black vertical line shows the median value, and the grey outline shows the 68\% confidence interval, or where the cumulative distribution function (CDF)'s value ranges from 0.16 to 0.84.
	(Top row) In both TT simulations, the temperature field has a mean value at $T = 0.5$ and a symmetric distribution around that peak with extrema at the fixed values of the boundaries. 
	(Bottom left) At early times in the FT simulation, the modal value of the PDF constantly moves left (towards the cold fixed-temperature boundary).
	(Bottom right) At late times, the temperature PDF from the cold fixed-temperature value (on the left) to the modal value is indistinguishable from the TT PDF, but from the mode to the fixed-flux boundary there is a large tail characterized by low-probability, hot elements.
	\label{fig:rbc_evolution_dynamics} }
\end{figure}

In Fig.~\ref{fig:rbc_evolution_dynamics}, we compare the time evolution of the temperature field of a classic-IC FT simulation with Ra$_{\partial_z T}$ = 4.83$\,\times 10^{10}$ to two TT simulations (with Ra$_{\Delta T} = 10^{10}$ and Ra$_{\Delta T} = 10^9$, respectively).
As shown in the top four panels, we see the expected convective roll solution in both TT simulations (top row) and at early and late times in the FT simulation (bottom row).
Interestingly, we find highly asymmetrical dynamics at early times in the FT simulation (bottom left), in which the temperature anomaly in the cold plume is much greater than in the warm plume.
This excess cold material slowly fills the domain and mixes, reducing the temperature difference between the top and bottom plates from $\Delta T = -1$ to $\Delta T = -\text{Nu}^{-1}$ in the relaxed state.
In this relaxed state, the supply of warm fluid from the bottom plume and cold fluid from the top plume come into balance, and the FT dynamics (bottom right) are indistinguishable from the TT dynamics (top right).

In the bottom four panels, we examine these temperature fields statistically by displaying their probability distribution functions (PDFs).
To create these PDFs, we sample the full simulation temperature field once every evolved freefall time, $\tau_{\text{ff, ev}}$, over the span of 500$\tau_{\text{ff, ev}}$.
We interpolate the (unevenly spaced) vertical Chebyshev grid points onto an evenly spaced grid before histogramming the flow values into 200 bins and creating the PDFs.

We find that this statistical analysis of the simulations tells the same story as the dynamical images shown above.
The temperature field in both of the TT simulations (top row) is dominated by the modal temperature of 0.5 in the bulk; a smaller fraction of the domain is filled with equal portions of hotter/colder material (mostly contained in the plumes), and the temperature field is rigidly bounded by the fixed-temperature boundary values.
The story is more complex for the FT simulation.
At early times (lower left), the FT simulation is characterized by two features: an extreme tail (to the left) that characterizes the cold plume at the upper boundary, and a migrating modal temperature that shifts from the right (hotter) to the left (cooler) as cold material mixes in the interior.
At late times (lower right), the FT simulation's PDF is indistinguishable from the TT PDF between the cold fixed-temperature boundary and the modal value.
From the modal value towards warmer temperatures, we find that the hot fixed-flux boundary is capable of producing more extreme temperature events and results in a more extended PDF tail.
This long tail is explored further in section \ref{sec:asymmetries}.

\begin{figure}[t!]
\includegraphics[width=\textwidth]{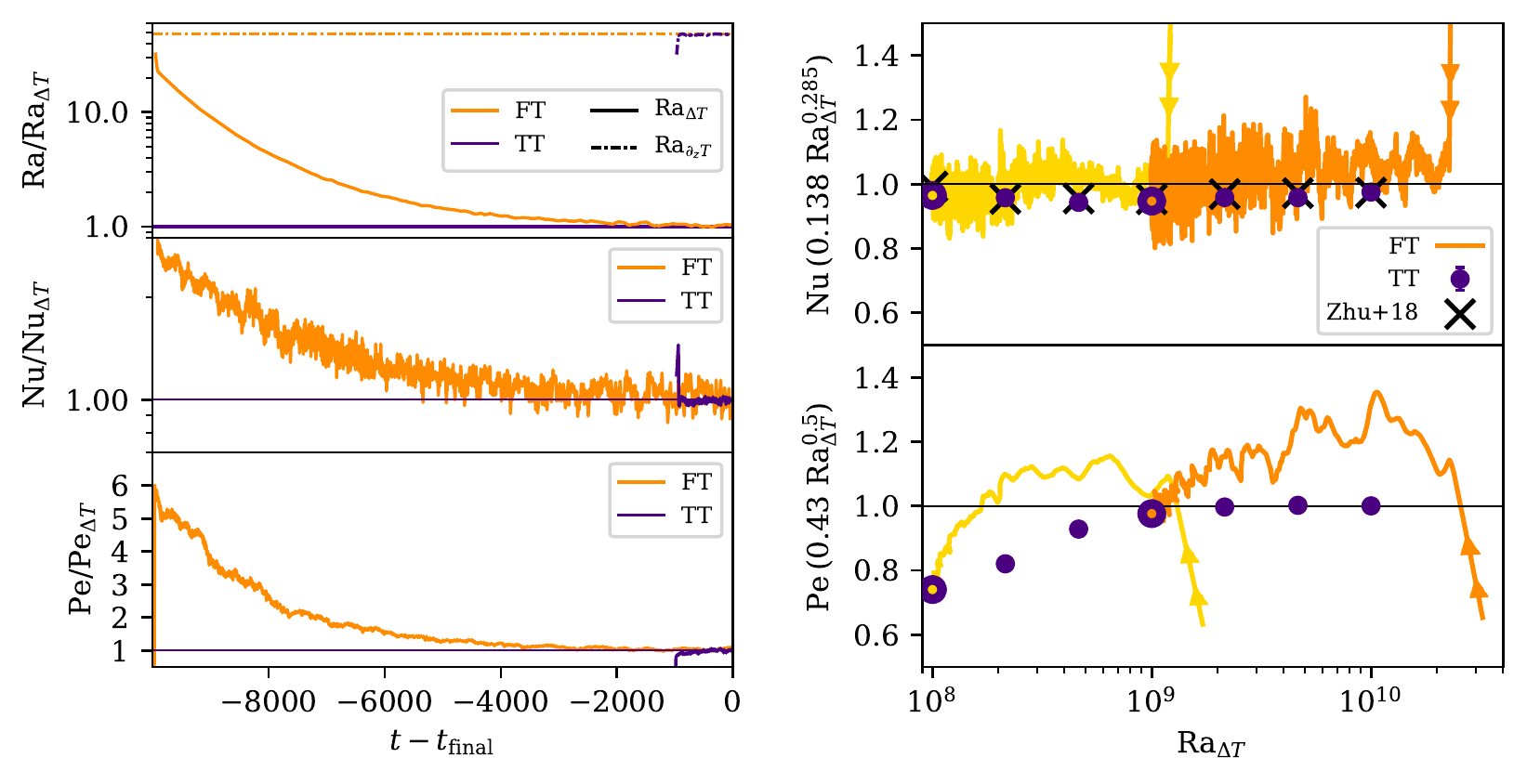}
\caption{ 
	(Left three panels) Time traces of scalar quantities in a classic-IC FT (orange, Ra$_{\partial_z T} = 4.81 \times 10^{10}$) and TT (purple, Ra$_{\Delta T} = 10^9$) simulation are shown.
	All traces have been averaged over a rolling window of 100 simulation time units to increase the clarity of the evolutionary trend.
	We display evolutionary traces of Ra (top, normalized by Ra$_{\Delta T}$ of the TT simulation) as well as Nu (middle) and Pe (bottom), both of which are normalized by their mean values measured over the last 500 freefall times of the TT simulation (reported in appendix \ref{app:table}).
	(Right two panels) Compensated scaling plots of Nu (upper) and Pe (lower) vs. Ra$_{\Delta T}$.
	Nu vs.~Ra is compensated by $(0.138 \text{Ra}_{\Delta T}^{0.285})$, the best-fit reported by ref.~\cite{johnston&doering2009}.
	Pe vs.~Ra is compensated by a Ra$_{\Delta T}^{1/2}$ law, the anticipated scaling of Pe \cite{ahlers&all2009}.
	The orange trace is the time evolution of the FT case from the left panels with the arrows showing the sense of time.
	The yellow trace shows the evolution of an FT case with Ra$_{\partial_z T} = 2.61 \times 10^9$.
	Purple circles are the measured values of Nu and Pe in our TT simulations (reported in appendix \ref{app:table}); error bars show the standard deviation of the sample mean and are smaller than the marker in all cases.
	The purple circles filled in with yellow and orange are the TT comparisons for the evolved states of the two FT cases.
	Black crosses show comparison TT simulations as reported by ref.~\cite{zhu&all2018}.
\label{fig:rbc_scalar_comparisons} }
\end{figure}

In the left panels of Fig.~\ref{fig:rbc_scalar_comparisons}, we examine the time evolution of scalar quantities from the FT simulation shown in Fig.~\ref{fig:rbc_evolution_dynamics} (orange lines) and compare it to the TT simulation with Ra$_{\Delta T} = 10^9$ (purples lines).
Simulation time is shown in nondimensional freefall units on the x-axis; the latest time displayed for each simulation, $t_{\text{final}}$, is subtracted for direct comparison of the relaxed states.
Traces of Ra$_{\Delta T}$ and Ra$_{\partial_z T}$ are shown in the top-left panel.
In the FT simulation, Ra$_{\Delta T}$ relaxes to its final value over thousands of simulation time units, and this final value is the input value of the equivalent TT case.
In comparison, Ra$_{\partial_z T}$ for the TT case instantaneously reaches its final value, which is the input value for the FT simulation.
This discrepancy in evolution timescales, where TT simulations evolve quickly and FT simulations evolve slowly, is also seen in the equilibration of Nu (middle panel) and Pe (bottom panel).

The right panels of Fig.~\ref{fig:rbc_scalar_comparisons} show that the relaxation of Ra$_{\Delta T}$ in FT simulations is akin to a sweep through Ra$_{\Delta T}$ parameter space.
The orange (Ra$_{\partial_z T} = 4.83 \times 10^{10}$, as on the left) and yellow (Ra$_{\partial_z T} = 2.61 \times 10^{9}$) lines show the evolution of FT simulations, and the arrows give the sense of time in the simulations.
For comparison, we plot results from TT simulations (purple circles) and the reported results of ref.~\cite{zhu&all2018} (black crosses).
The purple circles filled with orange and yellow circles are comparison TT simulations for the relaxed states of the FT simulations.
The top-right panel is a scaling plot for Nu vs.~Ra$_{\Delta T}$ compensated by the best fit reported in ref.~\cite{johnston&doering2009}.
The bottom-right panel is a scaling plot of Pe vs~Ra$_{\Delta T}$ compensated by the expected scaling \cite{ahlers&all2009}.
We find that FT simulations carry marginally more flux (higher Nu) and are more turbulent (higher Pe) than comparable TT simulations as they relax through this parameter space.
By binning Ra$_{\Delta T}(t)$, Nu$(t)$, and Pe$(t)$ into ten bins over the evolution of our FT simulations, we can quantify the path through parameter space that our FT simulations trace out.
By performing a least-squares fit to this data, the best-fit paths for Ra$_{\partial_z T} = 4.83 \times 10^{10}$ are $\text{Nu} = 0.0618\,\text{Ra}_{\Delta T}^{0.322}$ and $\text{Pe} = (5.73 \times 10^{-2})\,\text{Ra}_{\Delta T}^{0.597}$.
At Ra$_{\partial_z T} = 2.61 \times 10^{9}$, our best-fit paths are $\text{Nu} = 0.0845\,\text{Ra}_{\Delta T}^{0.310}$ and $\text{Pe} = (4.05 \times 10^{-3})\,\text{Ra}_{\Delta T}^{0.739}$.
By comparison, the best fit scaling laws for our TT cases are $\text{Nu} = 0.141\,\text{Ra}_{\Delta T}^{0.282}$ (very similar to the law reported in ref.~\cite{johnston&doering2009}), $\text{Pe} = (0.303)\,\text{Ra}_{\Delta T}^{0.516}$ (for Ra$_{\Delta T} \geq 10^9$), and $\text{Pe} = (2.41 \times 10^{-2})\,\text{Ra}_{\Delta T}^{0.64}$ (for Ra$_{\Delta T} < 10^9$).
These heightened values of Nu and Pe suggest that the dynamics do not immediately ``forget'' the higher-Ra$_{\Delta T}$ state that they recently timestepped through on their way to achieving thermal relaxation.

Achieving thermal relaxation in classic-IC FT simulations is computationally costly for two reasons: (1) the turbulent dynamics at the large initial Ra$_{\Delta T}$ require more spectral modes to resolve than the equilibrated state (compare the left and right dynamics in Fig.~\ref{fig:rbc_evolution_dynamics}), and (2) thousands of freefall times must pass during relaxation (see Fig.~\ref{fig:rbc_scalar_comparisons}).
For example, for the cases displayed in the left panels of Fig.~\ref{fig:rbc_scalar_comparisons} with a modest Ra$_{\Delta T} = 10^9$, the shown evolution of $10^4$ time units of the FT simulation cost $\sim 4.5 \times 10^5$ cpu-hours, while the TT equivalent case cost only $5.6 \times 10^4$ cpu-hours -- nearly an order of magnitude difference.
FT simulation dynamics evolve slowly during thermal relaxation, and these images, PDFs, and traces demonstrate the importance of waiting for thermal relaxation to be achieved when conducting an FT simulation.
In practice, in this work, we find that the thermal relaxation of FT simulations with classic ICs takes $\sim 2.5\sqrt{\text{Ra}_{\partial_z T} \text{Pr}}\,\text{Nu}^{-1}$ simulation freefall time units.

\newpage
\subsection{TT-to-FT \& Nu-based ICs: rapidly equilibrated FT simulations}
\label{sec:tt-to-ft}

\begin{figure}
\includegraphics[width=\textwidth]{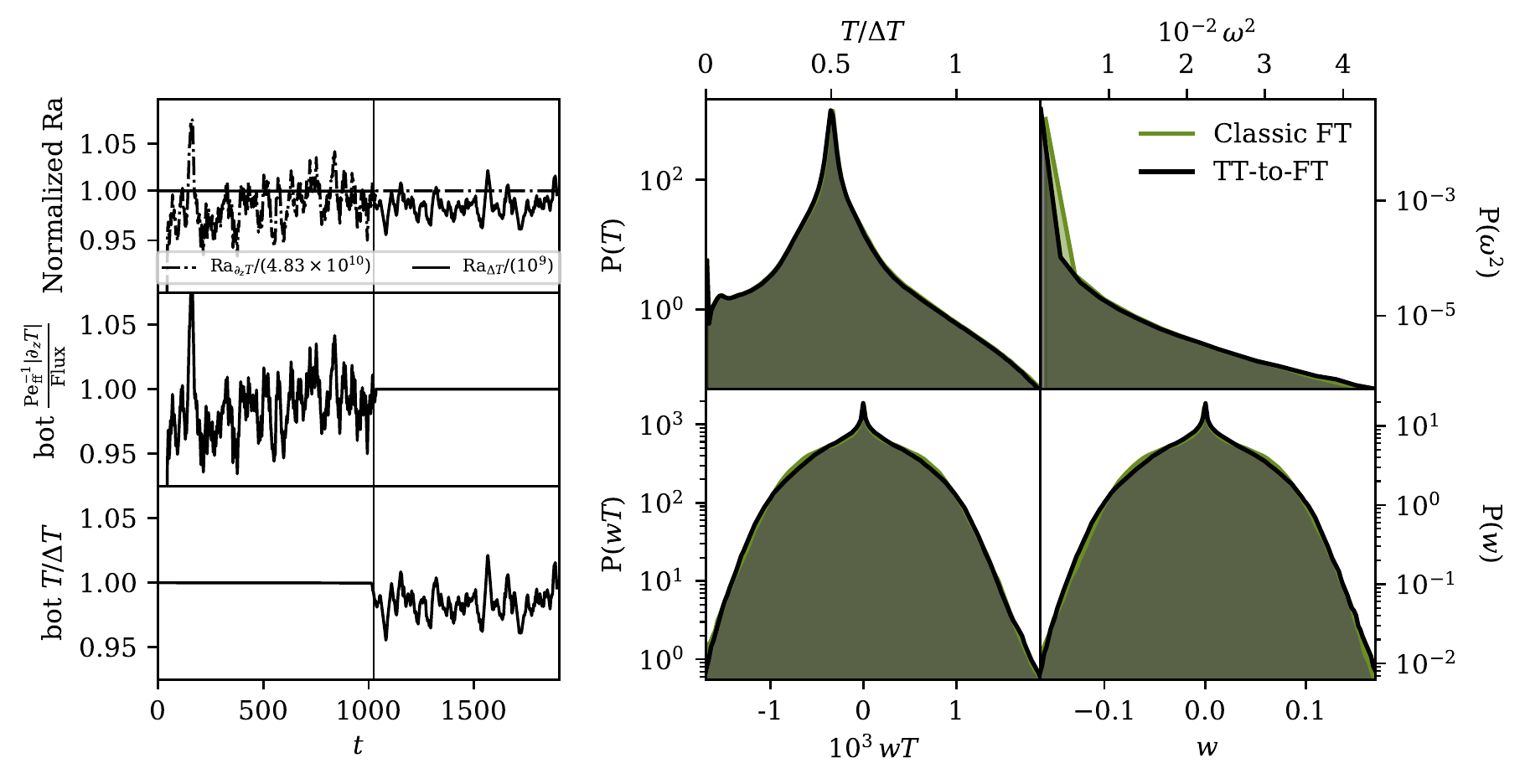}
\caption{ 
	(Left three panels) Time traces (in units of $\tau_{\text{ff, ev}}$) of scalar quantities, which have been averaged over a rolling time window of 25$\tau_{\text{ff, ev}}$, are shown for a simulation with Ra$_{\Delta T} = 10^9$ which starts with TT boundary conditions and then is switched to FT boundary conditions with Ra$_{\partial_z T} = 4.83 \times 10^{10}$.
	The time of the change of boundary conditions is denoted by the vertical black line.
	(Top panel) The evolution of Ra is shown; Ra$_{\partial_z T}/(4.83\times 10^{10})$ is shown as a dashed-dot line, while Ra$_{\Delta T}/10^9$ is shown as a solid line.
	The mean value of temperature gradient (middle panel) and temperature (bottom panel) at the bottom boundary are also shown.
	In the TT initial state, the temperature is held constant at a value of 1 and the temperature derivative fluctuates around a value of $\text{Nu}$.
	In the FT final state, the temperature derivative is held constant at a value of -1 and the temperature value fluctuates around a value of $\text{Nu}^{-1}$.
	(right four panels) PDFs are shown which compare TT-to-FT dynamics (black PDFs) to dynamics from the classic-IC FT case from Fig.~\ref{fig:rbc_scalar_comparisons} (green PDFs).
	We display the temperature field (upper left), enstrophy (upper right), nonlinear convective enthalpy flux (bottom left), and vertical velocity (bottom right).
\label{fig:rbc_restart_description} }
\vspace{-0.25cm}
\end{figure}

\subsubsection{TT-to-FT} Figs.~\ref{fig:rbc_evolution_dynamics} \& \ref{fig:rbc_scalar_comparisons} demonstrate that the statistically-stationary states of FT and TT simulations are similar in a qualitative sense and in their volume-averaged flow statistics.
It should therefore be possible to use results from a TT simulation to quickly reach the relaxed state of a comparable FT simulation, saving up to an order of magnitude in computational cost.
We show the results of using our TT-to-FT initial conditions procedure in practice in Fig.~\ref{fig:rbc_restart_description}.
In the left three panels, we display the temporal behavior of (top) Ra, (middle) the flux at the bottom boundary, and (bottom) the temperature difference between the top and bottom boundaries.
We take the full evolution of the Ra$_{\Delta T} = 10^9$ TT simulation shown in Fig.~\ref{fig:rbc_scalar_comparisons}, then change its boundary conditions to FT at Ra$_{\partial_z T} = 4.83\times 10^{10}$.
The change from TT to FT boundaries occurs at the time denoted by the thin vertical line.
Unlike in the FT case displayed in Fig.~\ref{fig:rbc_scalar_comparisons}, there is no thermal rundown in the FT state, due to the rapid relaxation achieved during the TT portion of the simulation.

In the right four panels of Fig.~\ref{fig:rbc_restart_description}, we compare PDFs of flow fields in this TT-to-FT simulation and the comparable classic FT simulation.
Shown are PDFs of the temperature field (upper left), enstrophy (upper right), convective flux (lower left), and vertical velocity (lower right).
In Table~\ref{table:pdf_values}, we display the first four moments of each of these distributions,
\begin{equation}
\begin{split}
&\mu(A) \equiv \sum_{i} A_i\,P(A_i)\,\Delta A,\qquad\qquad\qquad\qquad\qquad\qquad\,\,
\sigma(A) \equiv \sqrt{\sum_{i}[A_i-\mu(A)]^2 P(A_i) \Delta A},\\
&\mathcal{S}(A) \equiv \frac{1}{\sigma(A)^3}\sum_i [A_i-\mu(A)]^3 P(A_i) \Delta A,\qquad
\mathcal{K}(A) \equiv \frac{1}{\sigma(A)^4}\sum_i [A_i-\mu(A)]^4 P(A_i) \Delta A,
\end{split}
\label{eqn:pdf_moments}
\end{equation}
where $A$ is a flow quantity, $P(A)$ is the PDF of $A$, $\mu$ is the mean, $\sigma$ is the standard deviation, $\mathcal{S}$ is the skewness, $\mathcal{K}$ is the kurtosis, $\Delta A$ is the spacing between the discrete PDF bins, and $i$ is the index of the bin.
We specifically report the excess kurtosis, $\mathcal{K}_e = \mathcal{K}- 3$, to show how the $\mathcal{K}$ of our PDFs differs from the $\mathcal{K}$ of normal distributions.
The PDFs of all FT simulations agree well regardless of initial conditions, suggesting that all initial conditions do achieve a similar statistically stationary state.

We consider the classic-IC FT simulation to be in its evolved, relaxed state once a rolling temporal average of $\Delta T$ over 200$\tau_{\text{ff, ev}}$ converges to within 1\% of its final value.
We run our classic-IC FT simulations for at least $1000\tau_{\text{ff, ev}}$ past this point to ensure that the simulations are converged, and we gather statistics from the latest, most-converged simulation times that are available.
TT-to-FT and Nu-based IC FT simulations are only evolved for $1000\tau_{\text{ff, ev}}$, and we gather statistics from the second half of these simulation.

\subsubsection{Statistical Comparison of BCs and ICs}
From the moments of the PDFs presented in table \ref{table:pdf_values}, we conclude that FT and TT simulations are indistinguishable outside of their temperature fields.
The temperature PDF of the TT simulation unsurprisingly has a mean of 0.5, a $\mathcal{S}$ close to zero (no asymmetry between upflows and downflows), and an appreciable $\mathcal{K}_e$ (the tails, which primarily sample the plumes, are more important than they are in a normal distribution).
The FT temperature PDF, on the other hand, has a mean slightly larger than 0.5, a small but noticeable $\mathcal{S}$ (suggesting asymmetries between the F and T plates), and more $\mathcal{K}_e$ (implying more extreme plumes).
It is also interesting that, in all cases, the vertical velocity, $w$, and the vertical heat transport, $wT$, demonstrate PDFs whose tails are well-described by normal distributions.

The moments presented in table \ref{table:pdf_values} are not perfectly identical for all of the different FT cases, but we expect that this is due to the randomness of the turbulent motions in our simulations and our relatively short (500$\tau_{\text{ff, ev}}$) statistical sampling window.
If we were to run each of these simulations for multiple thermal diffusion times, and gather PDF statistics over thousands of freefall times, we would expect to find precisely the same distribution for each case.
However, each of the PDF moments for different FT cases agree to within a few percent (or are roughly zero), so we are satisfied that our modest sampling windows are sufficiently long to meaningfully compare our various simulations.

\begin{table}[t!]
\vspace{-0.5cm}
\caption{ 
	The first four moments, as defined in Eqn.~\ref{eqn:pdf_moments}, of each of the PDFs shown in Fig.~\ref{fig:rbc_restart_description} are displayed below for the TT case with Ra$_{\Delta T} = 10^9$ and all FT cases with Ra$_{\partial_z T} = 4.83\times 10^{10}$.
}
\vspace{-0.25cm}
\setlength{\tabcolsep}{12pt}
\label{table:pdf_values}
\begin{center}
\begin{tabular}{c c c c c c c}
\hline																	
Quantity 				&	BCs	& ICs	    &	$\mu$					&	$\sigma$				&	$\mathcal{S}$	&	$\mathcal{K}_e$ \\
\hline
$T/\Delta T$			&	TT	& Classic	&	$5.00 \times 10^{-1}$	&	$7.11 \times 10^{-2}$	&	$-1.53 \times 10^{-2}$	&	$1.71 \times 10^{1}$ \\
						&	FT	& Classic	&	$5.07 \times 10^{-1}$	&	$7.37 \times 10^{-2}$	&	$7.31 \times 10^{-1}$	&	$2.40 \times 10^{1}$ \\
						&	FT	& TT-to-FT	&	$5.08 \times 10^{-1}$	&	$7.34 \times 10^{-2}$	&	$7.18 \times 10^{-1}$	&	$2.40 \times 10^{1}$ \\
						&	FT	& Nu-based	&	$5.06 \times 10^{-1}$	&	$7.36 \times 10^{-2}$	&	$7.63 \times 10^{-1}$	&	$2.40 \times 10^{1}$ \\
\hline                                                                                                                      
$\omega^2/\Delta T$		&	TT	& Classic	&	$3.99 \times 10^{2}$	&	$4.73 \times 10^{2}$	&	$5.87 \times 10^{1}$	&	$6.68 \times 10^{3}$ \\
						&	FT	& Classic	&	$9.30 \times 10^{2}$	&	$4.79 \times 10^{2}$	&	$8.13 \times 10^{1}$	&	$1.76 \times 10^{4}$ \\
						&	FT	& TT-to-FT	&	$6.37 \times 10^{2}$	&	$5.02 \times 10^{2}$	&	$6.56 \times 10^{1}$	&	$8.69 \times 10^{3}$ \\
						&	FT	& Nu-based	&	$6.94 \times 10^{2}$	&	$4.84 \times 10^{2}$	&	$7.27 \times 10^{1}$	&	$1.20 \times 10^{4}$ \\
\hline                                                                                                                      
$wT/(\Delta T)^{3/2}$	&	TT	& Classic	&	$1.46 \times 10^{-3}$	&	$1.63 \times 10^{-1}$	&	$5.14 \times 10^{-2}$	&	$5.85  \times 10^{-2}$ \\
						&	FT	& Classic	&	$1.53 \times 10^{-3}$	&	$1.69 \times 10^{-1}$	&	$3.28 \times 10^{-2}$	&	$-8.19 \times 10^{-2}$ \\
						&	FT	& TT-to-FT	&	$1.51 \times 10^{-3}$	&	$1.67 \times 10^{-1}$	&	$3.63 \times 10^{-2}$	&	$1.29  \times 10^{-1}$ \\
						&	FT	& Nu-based	&	$1.50 \times 10^{-3}$	&	$1.70 \times 10^{-1}$	&	$4.37 \times 10^{-2}$	&	$1.95  \times 10^{-1}$ \\
\hline                                                                                                                    
$w/(\Delta T)^{1/2}$ 	&	TT	& Classic	&	$-3.11 \times 10^{-5}$	&	$3.25 \times 10^{-1}$	&	$1.64 \times 10^{-2}$	&	$-1.16 \times 10^{-2}$ \\
						&	FT	& Classic	&	$-3.80 \times 10^{-5}$	&	$3.33 \times 10^{-1}$	&	$1.75 \times 10^{-3}$	&	$-1.48 \times 10^{-1}$ \\
						&	FT	& TT-to-FT	&	$2.66  \times 10^{-5}$	&	$3.29 \times 10^{-1}$	&	$-1.21 \times 10^{-3}$	&	$7.89  \times 10^{-2}$ \\
						&	FT	& Nu-based	&	$2.41  \times 10^{-5}$	&	$3.35 \times 10^{-1}$	&	$1.08 \times 10^{-2}$	&	$1.37  \times 10^{-1}$ \\
\hline																	
\end{tabular}
\end{center}
\end{table}

\subsubsection{Nu-based ICs}
The time evolution of FT simulations with Nu-based ICs is similar to the time evolution of TT simulations with classic ICs, with a few small differences.
The interior temperature gradient ($\grad_{\text{interior}}$ in Eqn.~\ref{eqn:nu_based_gradT}) is slightly positive at high Ra, which means that the interior is marginally thermally stable while the boundary layers are thermally unstable.
During the convective transient, plumes from the boundaries eat away at this interior stratification over a few tens of freefall timescales, after which the interior is well mixed and a classic roll solution is achieved.
This behavior is quite different from classic ICs, in which the full domain is initially unstable and a roll solution is obtained immediately after the onset of nonlinear convection.
Regardless, the time required for the temperature field to reach statistical equilibrium is a few tens of freefall times rather than a few thousands of freefall times for classic ICs.
Per table \ref{table:pdf_values}, the evolved states of FT simulations with Nu-based ICs and classic ICs are very similar.

\subsection{Discussion of 2D results} 
We note briefly that Nu-based and TT-to-FT ICs are only two of many ways of accelerating the thermal relaxation of an FT simulation.
We discuss other mechanisms, and explore one in detail, in our previous work \cite{anders&all2018}.
We note however that the TT-to-FT setup described here is likely the least complicated mechanism for achieving rapid relaxation in a simplified RBC setup that we are aware of.
The successful degree with which this mechanism reproduces the evolved dynamics suggests that thermal relaxation occurs in two parts:
\begin{enumerate}
\item Changes to the simulation energy reservoir, and
\item Restratification of the experiment.
\end{enumerate}
The thermal energy reservoir of TT simulations does not change between the initial and final state.
The rapid relaxation of TT simulations, as well as FT simulations with Nu-based and TT-to-FT ICs, therefore suggests that experimental restratification occurs rapidly in RBC. 
The long rundown of classic-IC FT experiments on display in Fig.~\ref{fig:rbc_scalar_comparisons} is entirely due to the energy reservoir (the temperature jump across the domain) drifting over time.

As a final note, we find that measures of the velocity field (e.g., the kinetic energy and Pe) take a few hundred freefall timescales to relax to their final value in TT and FT (Nu-based IC) simulations in 2D at high values of Ra$_{\Delta T}\,\gtrsim 10^{8.67}$.
This velocity field relaxation happens despite instant \emph{thermal} relaxation of the temperature field for these simulations.
We find that the kinetic energy increases by less than a factor of two from its initial post-transient value to its final value in the statistically-stationary state.

\begin{figure}[p!]
\includegraphics[width=\textwidth]{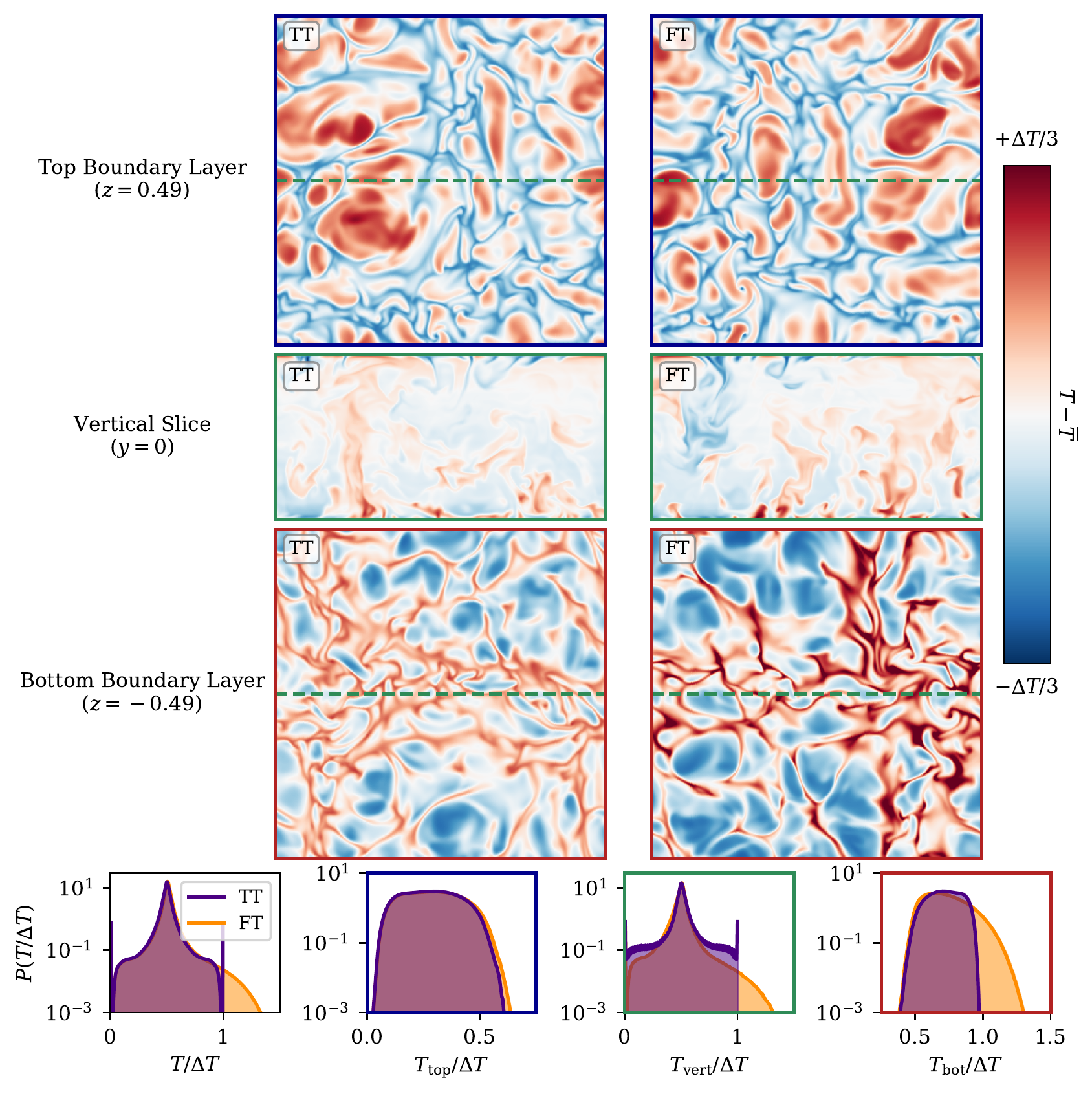}
\caption{ 
	(top six panels) Snapshots of the temperature anomaly are shown for select slices through the 3D TT simulation at Ra$_{\Delta T} = 10^8$ (left column) and the TT-to-FT simulation at Ra$_{\partial_z T} = 3.1 \times 10^9$ (right column).
	The top and bottom rows respectively show horizontal slices 1\% of the domain depth away from the top and bottom boundaries, while the middle row shows a vertical slice at $y = 0$.
	The $y = 0$ vertical slice intersects the two horizontal slices at the location indicated by the dashed green line.
	Visually, TT and FT dynamics are indistinguishable aside from the warm upflows near the bottom boundary.
	\\
	(Bottom four panels) PDFs of the temperature field are shown.
	From left to right, we display PDFs for the full 3D domain, the upper boundary layer slice at $z = 0.49$, the vertical slice at $y = 0$, and the bottom boundary layer slice at $z = -0.49$.
	The extreme temperature events which stand out visually near the bottom boundary layer can be clearly seen in all PDFs except for the one near the upper boundary.
	\label{fig:rbc_3D_panels} }
\end{figure}

\subsection{3D verification of 2D results}
It would be prohibatively expensive to timestep through the thermal relaxation of a 3D classic-IC FT simulation, even for our least turbulent Ra$_{\Delta T} \approx 10^8$ case.
We anticipate that the transient state would require a spectral coefficient resolution of $\gtrsim 512^3$ for adequate resolution, a factor of 64 times more coefficients than our comparable 2D case at 2048x1024.
This 2D case cost $1.21 \times 10^5$ cpu-hours, so we estimate that comparable 3D case would cost O(10 million) cpu-hours.
We found in previous work \cite{anders&all2018} at lower Ra$_{\partial_z T}$ that 3D classic-IC FT cases exhibited the same thermal rundown as 2D cases.
Here, we will focus on 3D comparisons of FT and TT simulations in the statistically-stationary state through the use of TT-to-FT initial conditions.

We find that equilibrated FT and TT simulations are analogous in a volume-averaged sense: measurements of Nu (31.4 for TT, 31.9 for FT) and Pe ($1.78 \times 10^3$ for both) are nearly indistinguishable between the two cases.
As in 2D, we find no difference between the PDFs of evolved quantities like the nonlinear transport ($wT$) between TT and FT simulations, but we do find differences between the evolved temperature fields ($T$).
In Fig.~\ref{fig:rbc_3D_panels}, we show dynamical slices of $T$ from 3D TT and FT simulations.
Near the top boundary and in the interior, the TT and FT dynamics are quite similar.
However, near the bottom boundary, the warm upflows in FT simulations are hotter relative to their counterparts in TT simulations.

At the bottom of Fig.~\ref{fig:rbc_3D_panels}, we compare PDFs of the full temperature field to PDFs of the temperature field in each of the shown slices (near the top boundary layer, a vertical slice of the interior, and near the bottom boundary layer).
The PDF of the temperature field over the full volume of these simulations shows the same features as in 2D (see Fig.~\ref{fig:rbc_evolution_dynamics}).
Interestingly, we find that within the upper boundary layer, the two cases have indistinguishable temperature fields, with temperature events ranging from the fixed boundary temperature ($T = 0$) to the temperature achieved in the interior ($T = 0.5\Delta T$) occuring with equal probability.
However, near the bottom boundary, the FT case instead exhibits a slightly larger number of neutral events ($T = 0.5\Delta T$) in addition to extreme temperature events ($T \geq \Delta T$).
We find that the PDF of the temperature field of a vertical slice through the domain at $y = 0$ is nearly identical to the PDF of the full-domain dynamics for the FT simulation, but that there are noticeable differences between these two PDFs for the TT simulation.
The differences between the full-volume and $y = 0$ PDFs in the TT case are due to the fact that our sampling window occurs while plume-launching sites are clustered around $y = 0$ more frequently than one would expect from a random distribution of plume-launching sites.
Over an infinite amount of time, these two PDFs should converge, but this demonstrates that it is important to be cautious when taking statistics from a 2D slice of a 3D simulation.

\FloatBarrier

\section{Results: Evolved structure, dynamics, and asymmetries in FT simulations}
\label{sec:results_dynamics}

\subsection{Evolved Structure}
\begin{figure}
\includegraphics[width=\textwidth]{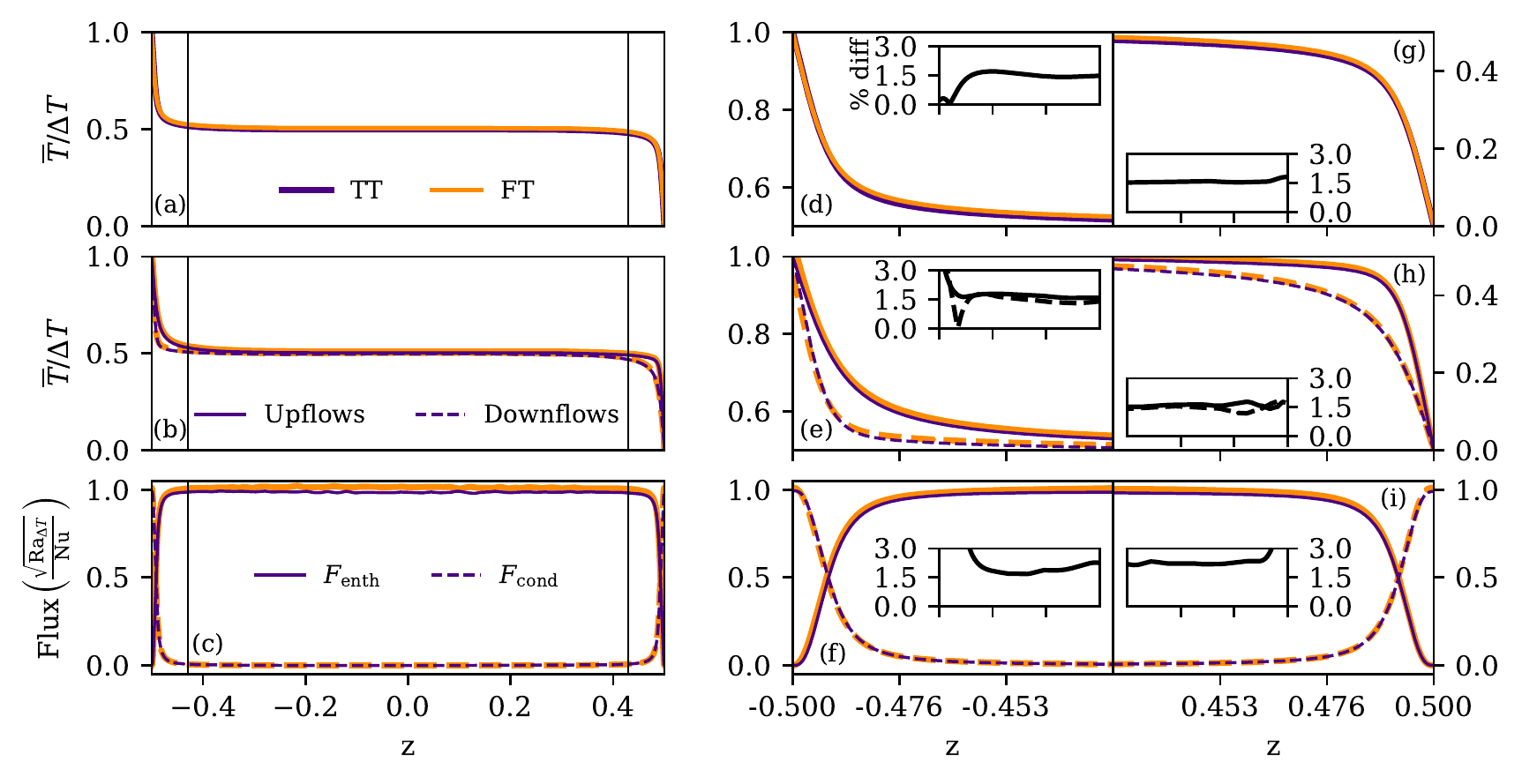}
\caption{ 
	(a-c) We compare time- and horizontally-averaged profiles from an FT (orange) and TT (purple) simulation at Ra$_{\Delta T} = 10^{9}$.
	Shown are the (a) temperature, (b) temperature in upflows (solid) and downflows (dashed), and (c) enthalpy (solid) and conductive (dashed) fluxes.
	The boundary layer regions are separated from the bulk by thin vertical lines and are examined in more detail in the right six panels (d-i).
	Displayed are the (d-f) bottom boundary layers and the (g-i) top boundary layers.
	The insets show the \% difference between the FT and TT solutions.
	There are slight (a few \%) differences between the two cases near the bottom boundary, but otherwise the two cases are nearly the same to within $\sim 1$\%.
\label{fig:rbc_1D_profiles} }
\vspace{0.5cm}
\end{figure}

In Fig.~\ref{fig:rbc_1D_profiles}, we compare the time- and horizontally-averaged profiles of the temperature and fluxes in the evolved FT and TT cases presented in Fig.~\ref{fig:rbc_scalar_comparisons}.
Time averages are taken over 500$\tau_{\text{ff, ev}}$, sampled once every 0.1$\tau_{\text{ff, ev}}$.
In the three left panels, we display profiles of (top) the mean temperature, (middle) the mean temperature in upflows (solid) and downflows (dashed), and (bottom) the convective enthalpy flux ($F_{\text{enth}} = wT$, solid) and the vertical conductive flux ($F_{\text{cond}} = -\Peff^{-1}\grad T$, dashed).
Most of the interesting structure is in the boundary layers, located between the sides of the plots and the thin vertical black lines.
Zoomed-in plots of the bottom and top boundary layers are respectively shown in the middle and right columns.
Inset panels show the percentage difference between the FT and TT solutions.
In the flux panels (bottom row), we do not plot the percentage difference in the conductive flux, as this quantity is undefined in the bulk of the interior where that flux is zero.
The conductive flux of the two cases agrees to within a few \% in the boundary layers, and the FT and TT cases differ by no more than 0.0025 in the plotted units in the interior. 
Here, we define the boundary layers as the heights above or below which conduction carries 95\% of the flux.
By this definition, the boundary layer depth is $\sim 0.024$ at Ra$_{\Delta T} = 10^9$, and we show three times this depth in the zoomed-in panels.

We find good ($\sim 1\%$) agreement between the FT and TT temperature profiles and enthalpy fluxes throughout the full depth of the domain, with slightly larger differences near the bottom boundary where the boundary conditions differ.
When we split the temperature profile into upflows and downflows, we find that FT upflows/downflows are slightly warmer/cooler than their TT counterparts at the hot, bottom boundary.
These (rather interesting) differences are illustrated in Fig.~\ref{fig:rbc_3D_panels}, and are explored further in the next section (also see Fig.~\ref{fig:rbc_dynamics_asymmetries}).
However, these differences do vanish in the interior and do not seem to affect the convective dynamics appreciably.

\subsection{Asymmetries induced by mixed boundary conditions}
\label{sec:asymmetries}

\begin{figure}
\includegraphics[width=\textwidth]{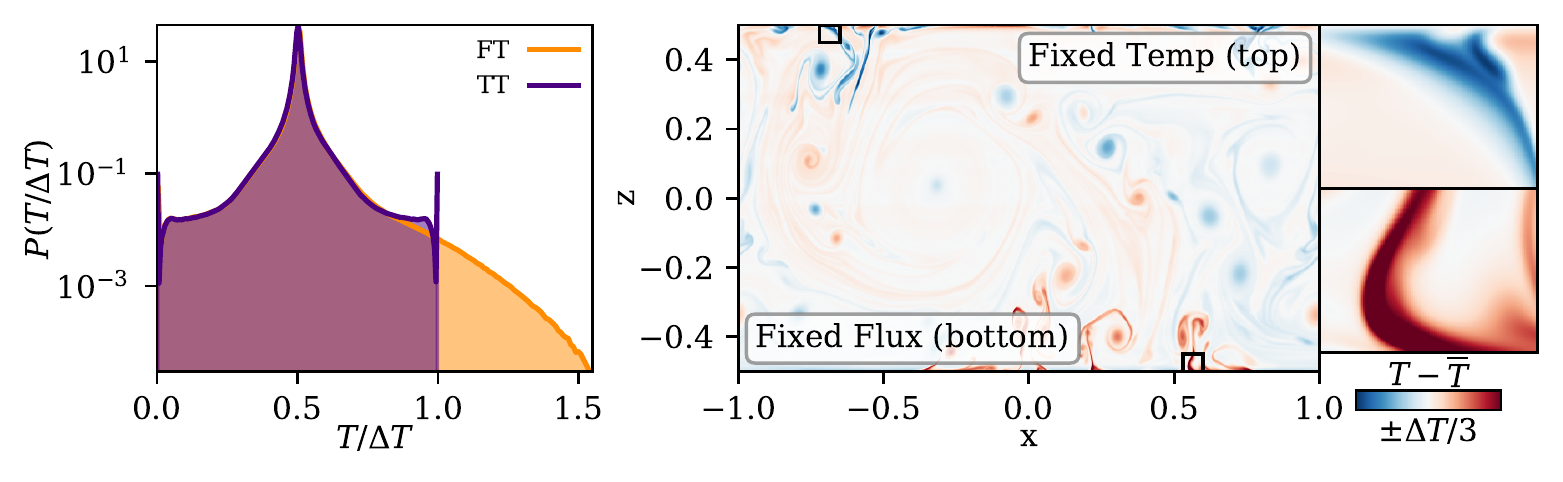}
\caption{ 
	(Left panel) PDFs of temperature measurements of a TT-to-FT (orange, Ra$_{\partial_z T} = 9.51 \times 10^{11}$) and TT (purple, Ra$_{\Delta T} = 10^{10}$) simulation are displayed.
	The right tail of the distribution (near the hot fixed-flux boundary for the FT case) shows that fixed flux boundaries achieve more extreme temperature events than fixed temperature boundaries.
	(Middle panel) A snapshot of the temperature anomaly in the FT simulation.
	Zoomed in views of the regions outlined in black boxes are shown in the right two panels.
	Near the top (fixed temperature) boundary, the temperature anomaly at the root of the plume vanishes, but this does not happen near the bottom (fixed flux) boundary, allowing for more extreme instantaneous values.
\label{fig:rbc_dynamics_asymmetries} }
\end{figure}

We now study in more detail the asymmetries introduced into a solution by FT boundaries.
We run a TT and TT-to-FT simulation at Ra$_{\Delta T} = 10^{10}$ and Ra$_{\partial_z T} = 9.51 \times 10^{11}$, respectively.
In Fig.~\ref{fig:rbc_dynamics_asymmetries}, we examine the dynamical nature of the asymmetries that the FT boundaries introduce into the simulation near the fixed-flux boundary.
In the left panel, we plot PDFs of the temperature fields of the two cases.
These PDFs agree remarkably well near the mean and for cold temperatures (near the fixed temperature boundary), but diverge in the tail of the PDF for hot temperatures where $T/\Delta T \gtrsim 0.8$, where the boundary conditions differ.
Interestingly, there are no temperature fluctuations which exceed the specified boundary values in the convective domain for TT simulations.
However, the FT PDF has a much longer tail and the FT solution achieves fluid parcels which are hotter than the average bottom boundary value by more than 50\%.
In order to understand how this is possible, we examine a snapshot of the FT simulation's temperature anomaly in the middle panel.
We have outlined a portion of a cold plume near the upper (fixed-temperature) boundary and a portion of a hot plume near the lower (fixed-flux) boundary, and these regions are magnified in the rightmost panels.
The fixed-temperature upper boundary suppresses temperature anomaly at the upper boundary and regulates the temperature minima which can be achieved.
The fixed-flux lower boundary does no such suppression and allows for extreme temperature values to be achieved in the plume-launching area, thus allowing for the asymmetry in the tails of the temperature PDF.

We note briefly that these asymmetries do not seem to affect mean or volume-averaged quantities in these simulations appreciably (see the agreement between FT and TT in Figs.~\ref{fig:rbc_scalar_comparisons} \& \ref{fig:rbc_1D_profiles}).
However, the fact that fixed-flux boundaries produce a wider temperature distribution with more extreme values may be important in some astrophysical studies.
We explore this further in the discussion in section \ref{sec:discussion}.

\section{Results: Rotating \RB Convection}
\label{sec:results_rotating}

Rotating convection is an excellent testbed for our TT-to-FT method.
In rotating convection, as Ra increases at a fixed value of the Ekman number (Ek), flows transition from the rotationally constrained to the rotationally unconstrained regime.
The scaling of Nu~vs.~Ra changes drastically between these two regimes, and is some blend of the two in the intermediary, marginally constrained regime.
Furthermore, the precise scaling law attained in the rotationally constrained regime differs as Ek changes, and these scaling laws are less straightforward and well-understood than their non-rotationally-constrained counterparts \cite{king&all2009, schmitz&tilgner2009, zhong&ahlers2010, julien&all2012, stevens&all2013, ecke2015, grooms2015, grooms&whitehead2015,  plumley&julien2019}.
As a result, the powerlaw used for Nu-based ICs would have to be a complex function of Ek, Ra, and boundary conditions for rotating RBC.
However, TT-to-FT should work generally for all parameters, so long as a TT simulation can be performed.
Here we explore the parameter space that the classic-IC FT simulation explores as it thermally relaxes and demonstrate that the TT-to-FT approach leads to rapid convergence for a rotating simulation.

We now study 3D rotating RBC with Ek$ \,= 10^{-6}$.
These simulations employ stress free boundary conditions which allow for the generation of mean flows such as large-scale vortices (LSV) \citep{stellmach&all2014, rubio&all2014, guervilly&all2014, guervilly&hughes2017, favier&all2014, favier&all2019, couston&all2020}.
We study a TT case at Ra$_{\Delta T} = 2.75\times 10^9$, and a classic-IC FT case at $\text{Ra}_{\partial_z T} = 2.1 \times 10^{10}$ (the supercriticality of the TT case is $\sim 3$).
We then take the TT case, do a TT-to-FT simulation, and compare the results of the TT-to-FT simulation to the classic-IC FT case.

\begin{figure}[t!]
\includegraphics[width=\textwidth]{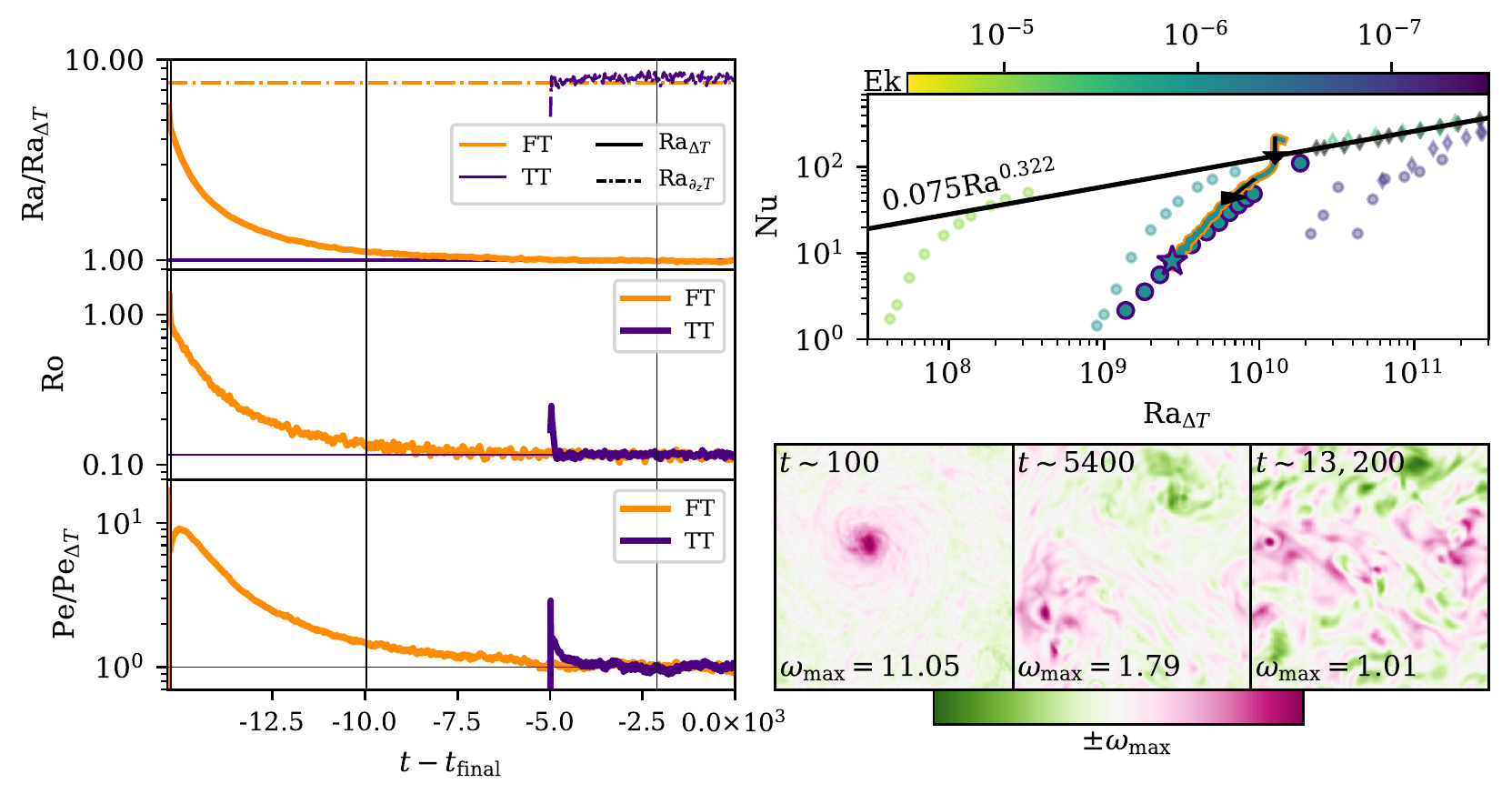}
\caption{ 
	(Left three panels) Time traces of scalar quantities in a classic-IC FT (orange, $\text{Ra}_{\partial_z T} = 2.1 \times 10^{10}$) and TT (purple, Ra$_{\Delta T} = 2.75\times 10^9$) simulation are shown.
	All traces have been averaged over a rolling window of 50 freefall time units to increase the clarity of the evolutionary trend.
	(Top panel) Ra, normalized by the input Ra$_{\Delta T}$ of the TT simulation.
	(Middle panel) Ro evolution of both simulations; the bulk flow of the FT simulation transitions from a marginally rotationally unconstrained state to a constrained state, while the TT simulation is always rotationally constrained.
	(Bottom panel) Pe evolution of the simulations is shown, normalized by the mean value measured over the last 500 freefall times of the TT simulation.
	(Upper right panel) Parameter space of Nu vs.~Ra$_{\Delta T}$ in rotating convection.
	Circular and diamond data points are respectively simulations and experimental data points from ref.~\cite{cheng&all2015}.
	The color of the data points signifies the Ekman number of the points, and grey diamonds are non-rotating.
	Data from our Ek$\,=\,10^{-6}$ FT experiment are shown as a thick orange line with a cyan interior, where the black arrows give the direction of time.
	Some TT simulations are shown as purple circles with a cyan interior, and the TT case which corresponds to the relaxed state of the FT simulation is shown as a star.
	(Bottom right panels) Snapshots of the vertically integrated z-component of the vorticity from the FT simulation.
	At early times (left panel), a powerful large-scale vortex with positive vorticity develops.
	This vortex slowly decays and becomes a vortex pair (middle panel), as seen in ref.~\cite{stellmach&all2014}.
	In the converged state, we see oscillatory behavior between this vortex pair behavior and jets (right panel).
	The TT case exhibits the oscillatory behavior between vortex pairs and jets throughout its whole evolution.
	The three vertical black lines in the left panels signify the times at which these snapshots are taken.
\label{fig:rotating_panels} }
\end{figure}

In the left three panels of Fig.~\ref{fig:rotating_panels}, we compare the time evolution of a classic-IC FT and TT case.
The top left panel shows the evolution of Ra$_{\partial_z T}$ and Ra$_{\Delta T}$.
Even in the presence of strong rotation, the TT simulation immediately equilibrates, but the FT case takes thousands of freefall times to achieve thermal relaxation.
In the middle panel, we show the evolution of Ro; the evolved flows in both simulations exhibit rotationally constrained dynamics with Ro $\,\approx 0.1$, but the flows in the FT simulation relax to this state from an initially unconstrained state (Ro $\,\approx 1$).
This implies that the thermal relaxation process can walk through the parameter space of flow balances (e.g., the balance between Inertial and Coriolis forces) in addition to the Ra$_{\Delta T}$ parameter space.
In the bottom panel, we display the evolution of Pe over time.
Strangely, the peak value of Pe occurs a few hundred freefall times after the convective transient.
After achieving this peak value, Pe monotonically decreases toward its relaxed state.
We find that the thermal relaxation of this case takes $\sim (2/3)\sqrt{\text{Ra}_{\partial_z T}\text{ Pr}}/\text{Nu}$ nondimensional freefall time units.

In the upper right panel of Fig.~\ref{fig:rotating_panels}, we plot Nu vs.~Ra for rotating simulations.
Select TT cases are plotted as cyan circles with purple outlines (where the cyan color denotes the value of Ek$\,=\,10^{-6}$ according to the color bar).
The evolution of the FT case in the left panels is shown as a thick orange line with a cyan interior and the black arrows show the direction of time.
The TT case that corresponds to the FT case is a purple star with a cyan interior.
We have additionally included some literature data from numerical simulations (circles) and experiments (diamonds) as reported in the appendix tables of ref.~\cite{cheng&all2015}.
These experiments were conducted in a cylindrical geometry at a different Pr, and are not meant to be one-to-one-comparable, but are meant to guide the eye to the nature of the parameter space of rotating convection.
The solid black line is the best-fit line for rotationally unconstrained simulations with Ra $\geq 10^{10}$ from ref.~\cite{cheng&all2015}.
As expected, the scaling of Nu vs.~Ra is steep in the rotationally constrained regime \cite{julien&all2012, plumley&julien2019}, which these simulations trace through.
We find that at values of Ra$_{\Delta T} \gtrsim 8 \times 10^9$, scaling laws start to flatten towards the unconstrained regime.
The FT simulation at all times when Ra$_{\Delta T} < 8 \times 10^9$ traces out $\text{Nu} = (1.08 \times 10^{-18})\,\text{Ra}_{\Delta T}^{2}$ and $\text{Pe} = (4.26 \times 10^{-20})\,\text{Ra}_{\Delta T}^{2.4}$.
The TT simulations with Ra$_{\Delta T} < 8 \times 10^9$ trace out $\text{Nu} = (9.36 \times 10^{-17})\,\text{Ra}_{\Delta T}^{1.79}$ and $\text{Pe} = (4.78 \times 10^{-19})\,\text{Ra}_{\Delta T}^{2.27}$.
As in Fig.~\ref{fig:rbc_scalar_comparisons}, the FT scaling laws are once again steeper than the comparable TT laws.

In the bottom right three panels of Fig.~\ref{fig:rotating_panels}, we plot the vertically integrated vertical vorticity in the simulation at three different times.
In the left panel, a dominant LSV which is aligned with the global rotation dominates the simulation at early times.
Over thousands of freefall times, this LSV evolves into a long-lived vortex pair, displayed in the middle panel.
Finally, in the evolved state, this vortex pair solution begins to oscillate with domain-wide jets, such as those displayed in the right panel.
We find that the TT solution shows this oscillatory behavior between vortex pairs and jets immediately and throughout the full 5000 freefall timescales of evolution that we simulated.

We suspect that the strange behavior of Pe in the bottom left panel can be explained by the evolution of the dominant flow structures over time.
At early times, the initially large value of Ra$_{\Delta T}$ in the FT case drives the displayed dominant LSV.
This powerful driving injects energy into the LSV, causing Pe to grow.
As Ra$_{\Delta T}$ and convective driving decrease over time, the LSV saturates and then starts to wind down, leading to the ``bump'' in the Pe trace.

We performed a TT-to-FT simulation starting from the evolved state of the TT simulation, and its time evolution matched that seen in Fig.~\ref{fig:rbc_restart_description}.
There was no long thermal evolutionary timescale.
In the equilibrated classic-IC FT simulation, we measured Nu$=7.90 \pm 0.01$, Pe$=1.70 \times 10^3$, and Ro$=0.115 \pm 0.006$.
In the TT-to-FT simulation, we measured Nu$=7.83 \pm 0.01$, Pe$=1.65 \times 10^3$, and Ro=$0.114 \pm 0.005$.
These Ro measurements are indistinguishable, Nu measurements are within 1\% of each other, and Pe measurements are within 3\% of each other, representing excellent agreement.

The difference in computational cost between classic-IC FT simulations and the TT/TT-to-FT simulations are even more striking here than in our previous examples.
The TT simulation shown in the left panels of Fig.~\ref{fig:rotating_panels} only cost $2.2 \times 10^3$ cpu-hours to run 5000 freefall times.
The TT-to-FT simulation cost an additional $1.28 \times 10^3$ cpu-hours to run 1000$\tau_{\text{ff, ev}}$.
By comparison, the cost of the FT simulation shown in the same panels was \emph{roughly three orders of magnitude larger}---$2.3 \times 10^6$ cpu-hours.
The TT and TT-to-FT simulations had a coefficient resolution of $128^3$.
The FT simulation's initial resolution required to resolve the convective transient was $512\times384^2$ coefficients.
We reduced the resolution to $256\times384^2$ after 100 freefall times, and then later to $128\times384^2$ after $\sim3.3 \times 10^3$ freefall times.
At each of these times, we found that lowering the \emph{horizontal} coefficient resolution of the simulation did not reproduce the simulation solution with fidelity.
This suggests that small scale turbulent velocity structures---which are injected by the vigorous transient and perhaps associated with the LSV---are long lived throughout the thermal evolution of the simulation.


\section{Conclusions \& Discussion}
\label{sec:discussion}
In short, we find that simulations with mixed thermal boundary conditions can experience a long thermal relaxation which is not experienced by simulations with two fixed-temperature boundaries; furthermore, to first order, mixed thermal boundary conditions do not introduce important asymmetries into the solution.

In this paper, we have studied the time evolution of \RB convection (RBC) under two different formulations of the thermal boundary conditions: ``FT'' boundaries, where the flux is fixed at the bottom and temperature is fixed at the top, and ``TT'' boundaries, where temperature is fixed at the top and bottom.
In the case of FT boundaries, we studied three different sets of initial conditions and examined both the nature of thermal relaxation and the equilibrated, statistically stationary state.
Through studying this relaxation and the relaxed states, we come to the following conclusions:
\begin{enumerate}
\item Thermal relaxation in RBC has two components: (a) changes in the energy reservoir and (b) changes in the stratification.
We find that the long relaxation of classic-IC FT simulations is due to changes in the energy reservoir; this reservoir is roughly constant in TT simulations and FT simulations with Nu-based or TT-to-FT ICs due to the lack of evolution of the temperature difference between the boundaries.
The rapid evolution of all of our simulations other than the classic-IC FT simulations suggests that RBC thermally restratifies itself instantaneously.
\item Dynamical measurements taken during thermal relaxation may be misleading.
Dynamics during the relaxation are more turbulent than in the evolved state, and exhibit evolving flow balances in the equation of motion (as quantified by e.g., the Rossby number).
This is principly a concern in systems like classic-IC FT simulations which can have very long thermal relaxation timescales compared to dynamical times.
\item The thermal relaxation process of a classic-IC FT simulation performs a sweep through Ra$_{\Delta T}$ parameter space.
We find that convective heat transport (the Nusselt number) and turbulent velocities (the P\'{e}clet number) are elevated above classic scaling laws along these parameter space sweeps.
\item Great computational expense achieving thermal relaxation in an FT simulation can be avoided by using the evolved state of a TT simulation as a ``better'' set of initial conditions for an FT simulation, or by constructing an initial state which is characterized by a temperature difference similar to the evolved one.
\item Despite minor asymmetries near the boundaries, we find no meaningful difference between the mean state of FT and TT simulations.
\end{enumerate}

We now describe some lessons that should be applied from this work to astrophysical convection, and comment on some open areas of research.
Throughout this work, we have made the assumption that convection is only ``interesting'' in its final, fully equilibrated state.
In nature, convection is not always in an equilibrium state.
For example, in the late stages of the lifetimes of stars, some core burning regions have sufficiently short lifetimes that they likely do not come into thermal relaxation \citep{clarkson&all2018, andrassy&all2020}.
The use of classic-ICs with FT boundaries that we have here considered to be a ``bad'' choice may help in understanding these transient lifetime stages.
However, for most convective studies where the lifetime of the natural convective system is much larger than its Kelvin-Helmholtz timescale (the time it would take for an astrophysical body to radiate its full gravitational potential energy given its current luminosity \cite{carroll&ostlie}), it is essential to study relaxed convection, and our results point towards the importance of either choosing good initial conditions (TT or TT-to-FT simulations) or running simulations to thermal relaxation.

One question which our study of RBC is not able to address is: how long does it take for a complex convective system to restratify?
Our fully convective domains restratified instantaneously, but it is likely that mixed convective-and-stably-stratified domains \citep{brummell&all2002, kapyla&all2019, pratt&all2017, korre&all2019} should have regions that are not turbulently mixed by convection which could also have long relaxation timescales.
It would be extremely helpful for future studies to examine relaxational timescales in systems where the energy reservoir is fixed, but where convection does not effectively mix the whole domain.
Fortunately, clever techniques (e.g., as we explored in ref.~\cite{anders&all2018}) can likely be used to rapidly restratify atmospheres in such simulations.

RBC is fundamentally symmetrical, but many natural convective processes occur in density-stratified domains in which the symmetries of the problem are broken.
In the present study, we observed that flux boundaries produce more extreme thermodynamic events than temperature boundaries.
In studies of overshooting convection, it is possible that plumes produced by a flux boundary layer could launch further into a stable layer than plumes produced by a temperature boundary.
Some authors have aimed to quantify the nature of overshooting plumes from a convective region into a stable region \cite{pratt&all2017, korre&all2019}, and it is unclear if different choices of boundary conditions could change the observed distribution of overshooting plumes observed there.

Some of the most complex astrophysical convection experiments aim to understand self-consistently evolving magnetic dynamos in rotating, spherical, magnetohydrodynamical domains \cite{brown&all2010, yadav&all2016, strugarek&all2017, strugarek&all2018}.
These dynamo simulations involve large numbers of timesteps through many freefall timescales in order to study the generation and evolution of magnetic fields and mean flows.
We found in our classic-IC FT rotating simulation that the unrelaxed state generated a mean flow (a LSV, Fig.~\ref{fig:rotating_panels}) that was much more intense and large-scale than the eventual flows that developed in the relaxed state.
If we had terminated our FT rotating simulation too early, we would not have seen the eventual destruction of this LSV or the later oscillatory behavior between jets and vortex pairs.
Many dynamo simulations are performed in highly turbulent regimes at the cutting-edge of what is achievable using modern computational resources.
As a result, timestepping through thousands of freefall timescales is not possible in these simulations.
It is therefore crucial that dynamo simulations be set up in such a manner as to avoid large changes to the system's energy reservoir such as those that we observed and studied here.

In conclusion, we note that our results here should provide astrophysical convection simulations with reason for optimism.
Some problems that we encounter (e.g., long thermal rundown in classic-IC FT simulations) can be completely avoided through a careful understanding of the numerical system being solved.

\begin{acknowledgments}
We thank the anonymous referee who suggested the idea of Nu-based ICs, and whose careful and thoughtful comments greatly improved both the quality and clarity of this manuscript.
We thank Daniel Lecoanet, who first pointed out to us the importance of examining Ra$_{\Delta T}$ in FT simulations. 
We'd especially like to thank Jeff Oishi, who encouraged us to pursue the differences between early and late FT dynamics and who graciously saved the reader from reading ``mixedFT'' (instead of FT) and ``fixedT'' (instead of TT).
EHA acknowledges that this work was supported by NASA Headquarters under the NASA Earth and Space Science Fellowship Program -- Grant 80NSSC18K1199.
LK acknowledges support from the George Ellery Hale Post-Doctoral Fellowship and from NASA grant 80NSSC17K0008.
This work was additionally supported by NASA LWS grant NNX16AC92G and NASA SSW grant 80NSSC19K0026. 
Computations were conducted with support by the NASA High End Computing (HEC) Program through the NASA  Advanced Supercomputing (NAS) Division at Ames Research Center on Pleiades with allocation GID s1647.
\end{acknowledgments}

\appendix
\section{Table of Simulations}
\label{app:table}
Input and output information for the simulations in this work are shown in Table~\ref{table:simulations}.
The codes used to run these simulations can be found online in the repository of supplemental materials \cite{anders&all2020a_supp}.

\begin{table}[ht]
\caption{
	Input and output values from the simulations in this work are shown; all simulations have a Prandtl number of 1.
	Input quantities are the boundary conditions (BCs), initial conditions (ICs), Rayleigh number (Ra), coefficient resolution (nx$\times$ny$\times$nz, or horizontal $\times$ vertical), and the total simulation run time in freefall units $t_{\text{simulation}}$ and in cpu-hours.
	Output quantities are the Nusselt (Nu), P\'{e}clet (Pe), and Rossby (Ro) numbers.
	Reported values of Nu, Re, and Ro are the sample mean over the last 500 evolved freefall time units for all rotating cases and all 2D cases except the Ra$_{\partial_z T} = 9.51\times10^{11}$ TT-to-FT case, where samples were taken over 200 evolved freefall time units; samples were taken over 350 freefall time units for 3D non-rotating cases.
	Reported uncertainties are the standard deviation of the sample mean; when the uncertainty is not reported, it is smaller than the number of reported digits.
	The ``Nu comp'' values are comparison Nu values reported in ref.~\cite{zhu&all2018}.
	Resolutions marked by a $*$ show the initial, highest resolution utilized in the simulation.
	The 2D FT Ra = $4.83 \times 10^{10}$ simulation's resolution was changed to $1024\times2048$ about 500 freefall time units after transient.
	The rotating FT case's resolution was reduced to $384^2\times256$ about one hundred freefall time units after transient, and was further reduced to $384^2\times128$ about $3.3\times 10^3$ freefall times after transient.
}
\setlength{\tabcolsep}{8pt}
\label{table:simulations}
\begin{center}
\resizebox{\textwidth}{!}{
\begin{tabular}{c c c c c c c c c c}
\hline																	
BCs	&	ICs	&	Ra	&	nx$\times$ny$\times$nz	& $t_{\text{simulation}}$	&	cpu-hours 	&	Nu	&	Nu comp	&	Pe  & Ro \\
\hline \hline \multicolumn{6}{c}{\vspace{-0.2cm}}\\
\multicolumn{7}{c}{\vspace{0.1cm}Non-rotating Runs ($\Gamma = 2$, no-slip, Ek$\,=\,\infty$)} \\
\hline
TT			&	Classic		&	$1.00 \times 10^8$		&	384$^2$x256	&	482			&	$1.95 \times 10^5$	&	$31.4$			&	---		&	$1.78 \times 10^3$ & --- \\
FT			&	TT-to-FT	&	$3.10 \times 10^9$		&	384$^2$x256	&	2221		&	$1.95 \times 10^5$	&	$31.9$			&	---		&	$1.78 \times 10^3$ & --- \\
TT			&	Classic		&	$1.00 \times 10^8$		&	1024x512	&	1023		&	$5.57 \times 10^3$	&	$25.4 \pm 0.1$	&	26.1	&	$3.18 \times 10^3$ & --- \\
FT			&	Classic		&	$2.61 \times 10^9$		&	2048x1024	&	9410		&	$1.21 \times 10^5$	&	$25.3 \pm 0.2$	&	26.1	&	$3.31 \times 10^3$ & --- \\
FT			&	TT-to-FT	&	$2.61 \times 10^9$		&	1024x512	&	5040		&	$3.21 \times 10^3$	&	$26.0 \pm 0.1$	&	26.1	&	$3.17 \times 10^3$ & --- \\
TT			&	Classic		&	$2.15 \times 10^8$		&	1024x512	&	1023		&	$5.73 \times 10^3$	&	$31.3 \pm 0.2$	&	31.2	&	$5.17 \times 10^3$ & --- \\
TT			&	Classic		&	$4.64 \times 10^8$		&	2048x1024	&	1024		&	$4.66 \times 10^4$	&	$38.4 \pm 0.3$	&	38.9	&	$8.60 \times 10^3$ & --- \\
TT			&	Classic		&	$1.00 \times 10^9$		&	2048x1024	&	1023		&	$5.58 \times 10^4$	&	$48.0 \pm 0.4$	&	48.3	&	$1.33 \times 10^4$ & --- \\
FT			&	Classic		&	$4.83 \times 10^{10}$	&	4096x2048*	&	19702		&	$5.56 \times 10^5$	&	$48.8 \pm 0.4$	&	48.3	&	$1.36 \times 10^4$ & --- \\
FT			&	TT-to-FT	&	$4.83 \times 10^{10}$	&	2048x1024	&	6136		&	$3.64 \times 10^4$	&	$49.0 \pm 0.4$	&	48.3	&	$1.32 \times 10^4$ & --- \\
FT			&	Nu-based	&	$4.83 \times 10^{10}$	&	2048x1024	&	6877		&	$3.64 \times 10^4$	&	$49.0 \pm 0.4$	&	48.3	&	$1.34 \times 10^4$ & --- \\
TT			&	Classic		&	$2.15 \times 10^9$		&	2048x1024	&	1029		&	$6.38 \times 10^4$	&	$60.4 \pm 0.5$	&	61.1	&	$1.99 \times 10^4$ & --- \\
TT			&	Classic		&	$4.64 \times 10^9$		&	3072x1536	&	1024		&	$3.29 \times 10^5$	&	$75.2 \pm 0.6$	&	76.3	&	$2.94 \times 10^4$ & --- \\
TT			&	Classic		&	$1.00 \times 10^{10}$	&	4096x2048	&	1039		&	$7.79 \times 10^5$	&	$95.3 \pm 0.7$	&	95.1	&	$4.30 \times 10^4$ & --- \\
FT			&	TT-to-FT	&	$9.51 \times 10^{11}$	&	4096x2048	&	2142		&	$1.22 \times 10^5$	& 	$95.3 \pm 1.0$ 	&	95.1	&	$4.29 \times 10^4$ & --- \\
\hline																	
\multicolumn{7}{c}{\vspace{0.1cm}Rotating Runs ($\Gamma = 0.481$, stress-free, Ek = $10^{-6}$)} \\
\hline																	
TT	&	Classic		&	$1.38 \times 10^9$		&	64$^2\times$128 	&	2565		&	$2.98 \times 10^3$	&	$2.17$			&	---		&	$2.84 \times 10^2$  & $(3.38 \pm 0.17) \times 10^{-2}$ \\
TT	&	Classic		&	$1.83 \times 10^9$		&	64$^2\times$128 	&	2545		&	$3.54 \times 10^3$	&	$3.56$			&	---		&	$5.28 \times 10^2$  & $(5.67 \pm 0.33) \times 10^{-2}$ \\
TT	&	Classic		&	$2.29 \times 10^9$		&	128$^3$				&	2537		&	$1.08 \times 10^4$	&	$5.61$			&	---		&	$8.91 \times 10^2$  & $(8.56 \pm 0.44) \times 10^{-2}$ \\
TT	&	Classic		&	$2.75 \times 10^9$		&	128$^3$				&	5035		&	$2.2 \times 10^3$	&	$8.04 \pm 0.01$	&	---		&	$1.71 \times 10^3$  & $(1.17 \pm 0.06) \times 10^{-1}$ \\
FT	&	Classic		&	$2.1 \times 10^{10}$	&	384$^2\times$512*	&	13950		&	$2.3 \times 10^6$	&	$7.90 \pm 0.01$	&	---		&	$1.70 \times 10^3$  & $(1.15 \pm 0.06) \times 10^{-1}$ \\
FT	&	TT-to-FT	&	$2.1 \times 10^{10}$	&	128$^3$				&	2775		&	$1.28 \times 10^3$	&	$7.83 \pm 0.01$	&	---		&	$1.65 \times 10^3$  & $(1.14 \pm 0.05) \times 10^{-1}$ \\
TT	&	Classic		&	$3.67 \times 10^9$		&	128$^3$				&	2532		&	$1.55 \times 10^4$	&	$12.5$			&	---		&	$3.39 \times 10^3$  & $(1.74 \pm 0.08) \times 10^{-1}$ \\
TT	&	Classic		&	$4.58 \times 10^9$		&	128$^3$				&	2530		&	$1.69 \times 10^4$	&	$17.6$			&	---		&	$4.77 \times 10^3$  & $(2.35 \pm 0.08) \times 10^{-1}$ \\
TT	&	Classic		&	$5.50 \times 10^9$		&	192$^3$				&	2402		&	$7.35 \times 10^4$	&	$22.8$			&	---		&	$6.38 \times 10^3$  & $(2.96 \pm 0.11) \times 10^{-1}$ \\
TT	&	Classic		&	$6.42 \times 10^9$		&	192$^3$				&	2226		&	$7.35 \times 10^4$	&	$29.5$			&	---		&	$7.86 \times 10^3$  & $(3.65 \pm 0.16) \times 10^{-1}$ \\
TT	&	Classic		&	$7.33 \times 10^9$		&	256$^3$				&	1147		&	$1.47 \times 10^5$	&	$36.2$			&	---		&	$9.52 \times 10^3$  & $(4.33 \pm 0.17) \times 10^{-1}$ \\
TT	&	Classic		&	$8.25 \times 10^9$		&	256$^3$				&	1079		&	$1.47 \times 10^5$	&	$43.0$			&	---		&	$1.10 \times 10^4$  & $(5.01 \pm 0.20) \times 10^{-1}$ \\
TT	&	Classic		&	$9.17 \times 10^9$		&	256$^3$				&	1030		&	$1.47 \times 10^5$	&	$48.9$			&	---		&	$1.24 \times 10^4$  & $(5.63 \pm 0.26) \times 10^{-1}$ \\
TT	&	Classic		&	$1.834 \times 10^{10}$	&	256$^3$				&	971.9		&	$1.84 \times 10^5$	&	$111$			&	---		&	$2.21 \times 10^4$  & $1.18 \pm 0.09$ \\
\hline																	
\end{tabular}
}
\end{center}
\end{table}

\end{document}